\documentclass[11pt,a4paper]{article}

\usepackage{amssymb}
\usepackage{amsfonts}
\usepackage{amsmath}
\usepackage{graphicx}
\usepackage{bm}
\usepackage{caption}
\usepackage{upgreek}
\usepackage[pdftex,colorlinks,bookmarksopen,bookmarksnumbered,citecolor=blue,urlcolor=red]{hyperref}
\usepackage{subfigure}
\usepackage{xcolor}
\usepackage[authoryear,comma,sort&compress]{natbib}
\newcommand{\toVect}[1]{{\boldsymbol{#1}}}

\newcommand{\dd}{\text{\,}\mathrm{d}}

\newcommand{\x}{\toVect{x}}
\newcommand{\Displacement}{{\toVect{u}}}

\newcommand{\E}{\toVect{E}}

\newcommand{\Strain}{\toVect{\strain}}
\newcommand{\strain}{\varepsilon}
\newcommand{\cauchyStress}{\widehat{\sigma}}
\newcommand{\hyperStress}{\widetilde{\sigma}}
\newcommand{\electricDisp}{\widehat{D}}

\newcommand{\flexo}{\mu}

\newcommand{\elast}{C}
\newcommand{\strGr}{h}

\title{Topology optimization of flexoelectric metamaterials with apparent piezoelectricity}
\author
{F.~Greco$^1$, D.~Codony$^{2,1}$, H.~Mohammadi$^3$, S.~Fern\'{a}ndez-M\'{e}ndez$^1$, I.~Arias$^{1,3\ast}$\\
\\
\small{$^1$ Laboratori de C\`{a}lcul Num\`{e}ric (LaC\`{a}N), Universitat Polit\`{e}cnica de Catalunya (UPC),\vspace{-.4em}}
\\\small{Campus Nord UPC-C2, E-08034 Barcelona, Spain}
\\
\small{$^2$ College of Engineering, Georgia Institute of Technology, Atlanta, GA 30332, USA,}
\\
\small{$^3$ Centre Internacional de M{\`e}todes Num{\`e}rics en Enginyeria (CIMNE), 08034 Barcelona, Spain}
\\
\small{$^\ast$ Corresponding author; E-mail: irene.arias@upc.edu.}
}
\date{}

\begin{document}
\maketitle
\begin{abstract} 
The flexoelectric effect, coupling polarization and strain gradient as well as strain and electric field gradients, is universal to dielectrics, but, as compared to piezoelectricity, it is more difficult to harness as it requires field gradients and it is a small-scale effect. {This drawback} can be overcome by suitably designing multiscale metamaterials made of a non-piezoelectric base material but exhibiting apparent piezoelectricity. We develop a theoretical and computational framework to perform topology optimization of the representative volume element of such metamaterials by accurately modeling the governing equations of flexoelectricity using a Cartesian B-spline method, describing geometry with a level set, and resorting to genetic algorithms for optimization. We consider a multi-objective optimization problem where area fraction competes with {each one of the} four fundamental piezoelectric functionalities (stress/strain sensor/actuator). We computationally obtain Pareto fronts, and discuss the different geometries depending on the apparent piezoelectric coefficient being optimized. Our results show that optimal material architectures strongly depend on the specific functional property being optimized, and that, except for stress actuators, optimal structures are low-area-fraction lattices. In general, we find competitive estimations of apparent piezoelectricity as compared to reference materials such as quartz and PZT ceramics. This opens the possibility to design devices for sensing, actuation and energy harvesting from a much wider, cheaper and effective class of materials.

\end{abstract}

\section{Introduction}

The property of some materials to transduce electrical fields into mechanical deformations and vice versa, known as piezoelectric effect \citep{martin1972piezoelectricity}, is crucial for the design of a variety of devices, such as sensors \citep{tressler1998piezoelectric,gautschi2006piezoelectric}, actuators \citep{sinha2009piezoelectric,gao2020piezoelectric}, energy harvesters \citep{erturk2011piezoelectric,safaei2019review} and microelectromechanical systems \citep{muralt2008recent,smith2012pzt}.  
However, these material are in general lead-based, brittle and limited to be used in a specific temperature range \citep{haertling1999ferroelectric}.
Because of these limitations, there has been intense research over the last years to develop piezoelectricity in a broader class of lead-free materials  \citep{ saito2004lead, hong2016lead, rodel2009perspective}. 

Besides material design at a molecular scale, an alternative idea is to achieve apparent piezoelectricity in metamaterials or composites made of non-piezoelectric constituents by an appropriate geometry or combination of materials \citep{Fousek1999, sharma2007possibility}. 
This can be achieved through flexoelectricity \citep{wang2019flexoelectricity}, a coupling mechanism between strain gradient and electric polarization (direct flexoelectricity), or electric polarization gradient and strain (converse flexoelectricity).  Unlike piezoelectricity which is only present in non-centrosymmetric dielectrics, flexoelectricity is a property of all dielectric materials including crystals, polymers, biomaterials, liquid crystals, etc. \citep{zubko2013flexoelectric, doi:10.1002/adma.201203852, jiang2013flexoelectric, WANG2019100570}. 
As a result of a non-uniform deformation applied on a generic dielectric material, regardless of the symmetry of its microscopic structure,  strain gradients  break locally the spatial inversion symmetry inducing an electric response. {However, the} question of how to design a material such that local flexoelectric couplings result in significant apparent piezoelectricity is far from obvious.

\cite{mocci2021geometrically} proposed a class of geometrically polarized architected dielectrics with apparent piezoelectricity. By considering periodic metamaterials made of non-piezoelectric flexoelectric materials, {they} showed that (1) geometric polarization (lack of geometric centrosymmetry) of the representative volume element (RVE) {topology} and (2) small-scale geometric features subjected to bending are enough to achieve an apparent piezoelectric behavior similar to that of Quartz and lead zirconium titanate (PZT) materials. Several  designs were proposed out of physical reasoning and their geometric parameters varied, concluding that different piezoelectric coupling coefficients pertinent to different functionalities, e.g.~strain/stress sensors or actuators  \citep{ikeda1996fundamentals,mocci2021geometrically}, require different designs. Here, the goal is to develop a systematic framework for the topology optimization of the flexoelectric unit cell of 2D apparently-piezoelectric metamaterials. This kind of flexoelectric periodic structure is particularly important to potentially upscale the flexoelectric effect and integrate it in conventional device configurations for electromechanical conversion at a macro-scale.

To accomplish this, we consider generalized periodic boundary conditions on an RVE, an optimization procedure based on genetic algorithms (GA) \citep{tomassini1995survey} and optimize for the four main apparent piezoelectric coupling coefficients \citep{ikeda1996fundamentals,mocci2021geometrically} pertinent to different piezoelectric applications. We approximate the system of fourth-order partial differential equations (PDE) using a direct Galerkin approach based on smooth uniform Cartesian B-spline basis functions. To describe the {topology} of the RVE, we resort to the level-set method {within} an immersed boundary approach \citep{codony2019immersed}.  Alternative methods to solve the flexoelectric PDE include $\mathcal{C}^1$ triangular elements \citep{yvonnet2017numerical}, mixed formulations \citep{mao2016mixed, deng2017mixed}, smooth meshfree schemes \citep{abdollahi2014computational,zhuang2020meshfree}, conforming isogeometric formulations \citep{codony2020modeling,thai2018large, liu2019isogeometric,sharma2020geometry,do2019isogeometric} or a C0 interior penalty method \citep{Ventura2021,balcells2022c0}. The combination of immersed boundary  Cartesian B-splines and level sets achieves conceptual simplicity, robustness, accuracy and the ability to easily {modify} geometry during optimization. Gradient-based topology optimization with level sets requires smoothing algorithms and re-initializations to maintain the signed distance property  for the numerical convergence \citep{lopez2022isogeometric}.
Despite the higher computational cost of GA, their implementation is straightforward, can be trivially parallelized, and are better suited for design space exploration for global minima. They have been successfully used in the topology optimization of mechanical structures   \citep{jenkins1991towards,erbatur2000optimal, hajela1995genetic,coello2000multiobjective} and electromagnetic systems  \citep{im2003hybrid}.

Topology optimization has been applied previously to flexoelectric nanostructures \citep{nanthakumar2017topology,ghasemi2017level,ghasemi2018multi,zhang2022flexoelectric,lopez2022isogeometric,hamdia2019novel}, by means of level sets and mixed finite elements  \citep{nanthakumar2017topology}, isogeometric approximations and level sets \citep{ghasemi2017level,ghasemi2018multi}, a morphable void approach \citep{zhang2022flexoelectric}, a phase-field method with a diffuse boundary \citep{lopez2022isogeometric}, or a multi-field
micromorphic approach \citep{ortigosa2022computational}. 
Optimization algorithms include gradient methods, Monte Carlo algorithms \citep{hamdia2022multilevel}, as well as deep learning  approaches \citep{hamdia2019novel}. These works focused on finite structures (beams under bending and compressed blocks or pyramids) and optimized a ratio between mechanical and electric energy, rather than a direct measure of device performance for a given functionality. Interestingly, none of these approaches produce the kind of geometrically polarized microstructures proposed by \cite{mocci2021geometrically}. 
Instead, in the present work, the microstructural topology of architected materials is optimized for apparent piezoelectric performance and {some of} the resulting RVEs are reminiscent of those proposed by \cite{mocci2021geometrically}.
In a recent work by \cite{chen2021topology}, a similar philosophy is adopted with the opposite goal: architected materials made of piezoelectric composites in order to maximize their apparent flexoelectric behavior.

The outline of the paper is the following. We first {recall} the flexoelectric problem {statement} into the representative volume element, then we discuss the numerical discretization and the level set geometry description; we also discuss how GA are used and we {propose} a {methodology} to obtain {meaningful} material configurations; finally we show and discuss the results for the optimization of different apparent piezoelectric coefficients. Concluding remarks follow.


\section{Formulation for flexoelectric representative volume element\label{sec:2}}

\begin{figure}[b]\centering  	\includegraphics[width=.8\textwidth,clip,keepaspectratio,angle=0]{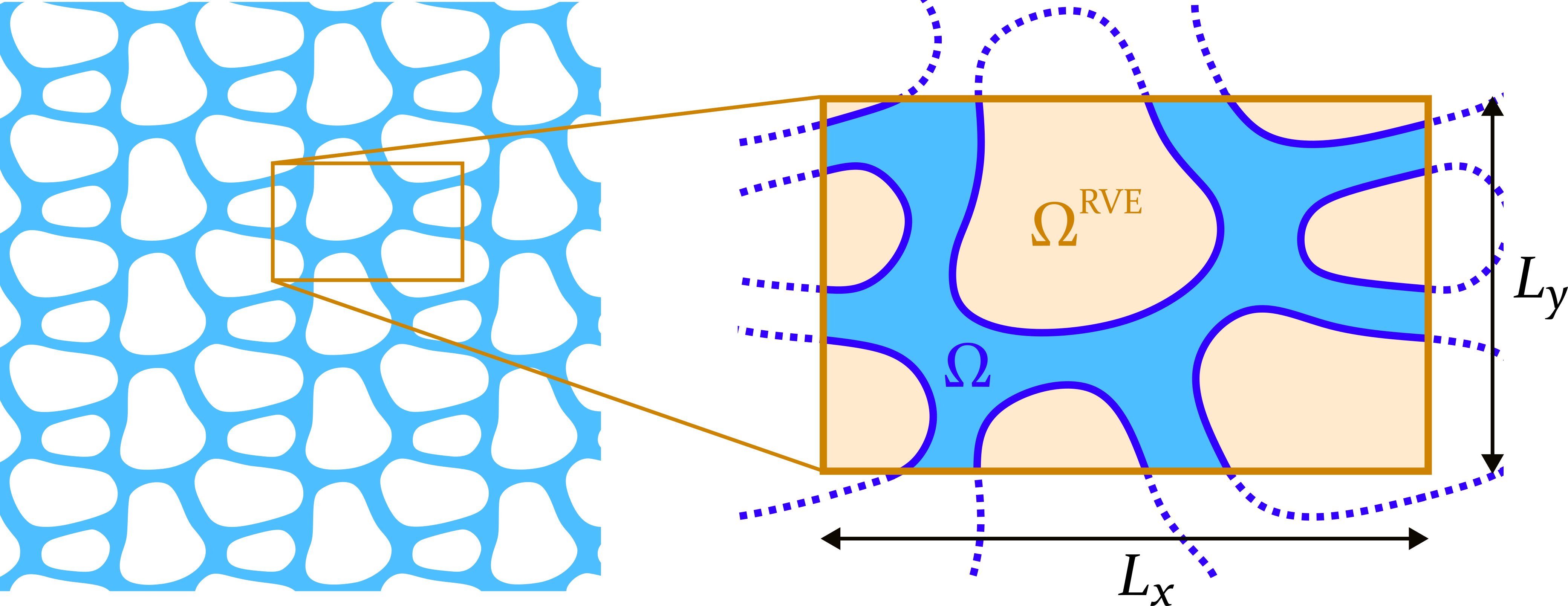}	
	\caption{Architected metamaterial (left) and representative volume element (right).} 
	\label{fig:RVE}
\end{figure}

We consider a periodic flexoelectric metamaterial, and choose the displacement field $\bm{u}(\x)$ and the electric potential $\phi(\x)$ as the primal variables characterizing {its state}. In the following, repeated indices imply sum over spatial dimensions, and indices after a comma $(,\!)$ denote the partial space derivative along {those directions}. We formulate the flexoelectric problem within the unit cell, referred to as representative volume element (RVE), c.f.~Fig.~\ref{fig:RVE}. 
Because the RVE is a repetitive unit in a periodic material, we need to consider macroscopic generalized periodicity conditions discussed below. This is not the case for most of the previous work on flexoelectric topology optimization  focusing on non-repetitive structures.

The strong form of the boundary value problem in $\Omega\subset\Omega^\textrm{RVE}=[0,L_x]\times[0,L_y]$ of a linear Lifshitz-invariant \citep{codony2021mathematical} flexoelectric metamaterial is described next. 
The constitutive relations for the linear flexoelectric material relate the stress tensor, electric displacement and their higher-order counterparts with the primal variables $\bm{u}$, $\phi$ and the material tensors as follows:
\begin{subequations}\label{elec_tensors}\begin{align}
\cauchyStress_{ij}(\Displacement,\phi)&=\elast_{ijkl}\strain_{kl}(\Displacement)
+\frac{1}{2}\flexo_{lijk}E_{l,k}(\phi),
\\
\hyperStress_{ijk}(\Displacement,\phi)&
=\strGr_{ijklmn}\strain_{lm,n}(\Displacement)-\frac{1}{2}\flexo_{lijk}E_{l}(\phi),
\\
\electricDisp_l(\Displacement,\phi)
&=\epsilon_{lm}E_{m}(\phi)
+\frac{1}{2}\flexo_{lijk}\strain_{ij,k}(\Displacement),
\\
\widetilde{D}_{kl}(\Displacement,\phi)
&
=M_{mnlk}E_{m,n}(\phi)-\frac{1}{2}\flexo_{lijk}\varepsilon_{ij}(\Displacement),
\end{align}\end{subequations}
where  $\strain_{ij}(\Displacement)=(u_{i,j}+u_{j,i})/2$ is the  strain field,  $E_l(\phi) = -\phi_{,l}$ the electric field,
$\elast_{ijkl}$ the elasticity tensor (in terms of Young modulus $Y$ and Poisson ratio $\nu$), $\strGr_{ijklmn}$ the strain-gradient elasticity tensor (in terms of $Y$, $\nu$ and the mechanical length scale $\ell_\textrm{mech}$), $\epsilon_{lm}$  the dielectricity tensor (in terms of the dielectric permittivity $\epsilon$), $M_{mnlk}$  the gradient dielectricity tensor (in terms of $\epsilon$ and the electrical length scale $\ell_\textrm{elec}$)
and $\flexo_{lijk}$ the flexoelectric tensor (in terms of the longitudinal $\mu_\textrm{L}$, transversal $\mu_\textrm{T}$ and shear $\mu_\textrm{S}$ flexoelectric coefficients). See details in Appendix A in~\cite{codony2021mathematical}. The  fourth-order Euler-Lagrange equations stating balance of linear momentum and Gauss's law for dielectrics are then
\begin{subequations}\label{eq_EulerLagrange}
\begin{align}
\cauchyStress_{ij,j}(\Displacement,\phi)-\hyperStress_{ijk,kj}(\Displacement,\phi)&=0 \qquad\text{in }\Omega,\\
\hfill\electricDisp_{l,l}(\Displacement,\phi)-\widetilde{D}_{lk,kl}(\Displacement,\phi)&=0\qquad \text{in }\Omega,
\end{align}
\end{subequations}
where we assume no volumetric external loads. 

The components of the primal unknowns are $\mathcal{C}^1$-continuous in $\Omega$ due to the high-order nature of the PDE system. The boundary of the domain $\partial\Omega$ has two parts, {one on} the boundary of the RVE $\partial\Omega^\text{RVE}$, which is not an actual material boundary and is subjected to generalized periodicity conditions, and the physical part of the boundary $\partial\Omega\setminus\partial\Omega^\text{RVE}$. Homogeneous Neumann conditions are considered at every physical boundary in the RVE, i.e.~there are no tractions $\bm{t}$, double tractions $\bm{r}$, edge forces $\bm{j}$, electric charges $w$, double electric charges $\mathfrak{r}$, or edge charges $\wp$ on the physical part of $\partial\Omega$. Mathematically,
    \begin{align}\label{eq:homNeumann}
        &\bm{t}=\bm{r}=\bm{j}=\bm{0},\quad w=\mathfrak{r}=\wp=0\qquad\text{on }\partial\Omega\setminus\partial\Omega^\text{RVE}.
    \end{align}
We refer to Ref.~\cite{codony2021mathematical} for a detailed description of the Neumann quantities involved in Eq.~\eqref{eq:homNeumann}.

A common assumption in boundary value problems of periodic metamaterials is that the primal variables are \emph{generalized-periodic} \citep{balcells2022c0,barceloinprep}, a procedure that allows us to consider macroscopic strains and electric fields at the level of the RVE. Generalized periodic conditions require that primal field gradients are periodic functions within the RVE, whereas the primal fields themselves (displacement and electric potential) are continuous across contiguous RVE modulo the  jumps resulting from the macroscopic strain and electric fields. For second order PDE, these conditions also ensure that the interfaces between contiguous RVE are in equilibrium \citep{barceloinprep}.

In the present paper, the fourth-order nature of the PDE requires the second gradient of the primal variables to be periodic as well, in order to ensure equilibrium at the RVE interfaces \citep{barceloinprep,barcelo2022weak}. Hence, the displacement field $\Displacement(\x)$ and the electric potential $\phi(\x)$ are {required} to be \emph{high-order} generalized-periodic functions, with $\mathcal{C}^1$-continuous components. They can be expressed as
\begin{subequations}
\label{eq:StateDecomposition}
    \begin{align}
        \Displacement(\x)&=\Displacement^\text{P}(\x)+ \overline{\mathbf{G}}\cdot \x, \\
        \phi(\x)&=\phi^\text{P}(\x)- \overline{\E}\cdot \x,
    \end{align}
\end{subequations}
where $\Displacement^\text{P}(\x)$ and $\phi^\text{P}(\x)$ are periodic functions in $\Omega^\textrm{RVE}$ representing the microscopic response of the material, whereas the quantities $\overline{\mathbf{G}}$ (a $2\times 2$ matrix) and $\overline{\E}$ (a vector in $\mathbb{R}^2$) represent the macroscopic displacement and electric potential gradients, and are constant over $\Omega^\textrm{RVE}$. The gradients of Eq.~\eqref{eq:StateDecomposition} are
\begin{subequations}
\label{eq:StateDecompositionGrad}
    \begin{align}
        \mathbf{G}(\x)\equiv \nabla\Displacement(\x)&=\nabla\Displacement^\text{P}(\x)+ \overline{\mathbf{G}}, \\
        \E(\x) \equiv -\nabla\phi(\x)&=-\nabla\phi^\text{P}(\x)+ \overline{\E},
    \end{align}
\end{subequations}
which are periodic functions in $\Omega^\textrm{RVE}$. The macroscopic fields $\overline{\mathbf{G}}$ and $\overline{\E}$ induce the following generalized-periodic conditions at the boundary of the RVE:
\begin{subequations}
\label{eq:StateDecomposition3}
    \begin{align}
        \Displacement(\xi=L_\xi)-\Displacement(\xi=0)&=\Displacement^\text{P}(\xi=L_\xi)+ \overline{\mathbf{G}}\cdot \boldsymbol{e}_\xi L_\xi - \Displacement^\text{P}(\xi=0) = \overline{\mathbf{G}}\cdot \boldsymbol{e}_\xi L_\xi, \\
        \phi(\xi=L_\xi)-\phi(\xi=0)&=\phi^\text{P}(\xi=L_\xi)- \overline{\E}\cdot \boldsymbol{e}_\xi L_\xi - \phi^\text{P}(\xi=0) = -\overline{\E}\cdot \boldsymbol{e}_\xi L_\xi,
    \end{align}
\end{subequations}
for $\xi=x,y$, and where $\boldsymbol{e}_\xi$ is the unit vector of the canonical basis along the $\xi-$direction; which imply
\begin{subequations}\begin{align}\label{eq:macroEPS}
    \overline{\mathbf{G}}(\Displacement)&=\left[
    \begin{array}{ccc}
         (\Displacement_x(x=L_x)-\Displacement_x(x=0))/L_{x} & 
         (\Displacement_y(x=L_x)-\Displacement_y(x=0))/L_{x}\\
         (\Displacement_x(y=L_y)-\Displacement_x(y=0))/L_{y} & 
         (\Displacement_y(y=L_y)-\Displacement_y(y=0))/L_{y}
    \end{array}\right],
\\\label{eq:macroEF}
    \overline{\E}(\phi)&=\left[
    \begin{array}{c}
         -\left(\phi(x=L_x)-\phi(x=0)\right)/L_{x} \\
         -\left(\phi(y=L_y)-\phi(y=0)\right)/L_{y} \\
    \end{array}\right].
\end{align}
\end{subequations}

Following Refs.~\cite{barceloinprep,barcelo2022weak}, the macroscopic gradient $\overline{\mathbf{G}}(\Displacement)$ is decomposed in its {symmetric and skew-symmetric components:}
\begin{subequations}\begin{align}
\overline{\Strain}(\Displacement)&=(\overline{\mathbf{G}}(\Displacement)+\overline{\mathbf{G}}(\Displacement)^\textrm{T})/2,\\
\overline{\mathbf{W}}(\Displacement)&=(\overline{\mathbf{G}}(\Displacement)-\overline{\mathbf{G}}(\Displacement)^\textrm{T})/2,
\end{align}\end{subequations}
where $\overline{\Strain}(\Displacement)$ is the macroscopic strain, and $\overline{\mathbf{W}}(\Displacement)$ is the macroscopic infinitesimal rotation (spin), which  corresponds to a rigid body motion and hence does not play any role in the Euler-Lagrange equations in \eqref{eq_EulerLagrange}. In the following, we choose to prevent {rigid body rotations} by applying the condition
\begin{align}
\overline{\mathbf{W}}(\Displacement)=\mathbf{0}
\qquad\leftrightarrow\qquad
\overline{{G}}_{xy}(\Displacement) = \overline{{G}}_{yx}(\Displacement)
\qquad\leftrightarrow\qquad
\overline{\mathbf{G}}(\Displacement) = \overline{\Strain}(\Displacement).
\end{align}
As shown in Refs.~\cite{barceloinprep,barcelo2022weak}, the weak formulation of the boundary value problem presented above{, prior to enforcing macroscopic conditions,} is
\begin{multline}
\label{eq:FinalWeakForm}
    \frac{1}{|\Omega^{\textnormal{RVE}}|}\int_{\Omega}\left(\cauchyStress_{ij}\!\left(\Displacement,\phi\right)\strain_{ij}\!\left(\delta\Displacement\right)-\electricDisp_l\!\left(\Displacement,\phi\right) E_l\!\left(\delta\phi\right)+\hyperStress_{ijk}\!\left(\Displacement,\phi\right)\strain_{ij,k}\!\left(\delta\Displacement\right)-\widetilde{D}_{lm}\!\left(\Displacement,\phi\right)E_{l,m}\!\left(\delta\phi\right)\right)\!\dd \Omega
    \\=
    \overline{\sigma}_{ij}\!\left(\Displacement,\phi\right) \overline{\strain}_{ij}\!\left(\delta\Displacement\right) - \overline{D}_l\!\left(\Displacement,\phi\right)\!\overline{E}_l\!\left(\delta\phi\right),
\end{multline}
for every admissible $\delta\Displacement(\x)$ and $\delta\phi(\x)$, where $|\Omega^{\textnormal{RVE}}|=L_xL_y$ is the total volume of the RVE (including material and void phases), and $\overline{\bm{\sigma}}$ and $\overline{\toVect{D}}$ are the macroscopic stress and macroscopic electric displacement. Intuitively, the left hand side of Eq.~\eqref{eq:FinalWeakForm} corresponds to the microscopic internal work density, and the right hand side is the work density at the macroscopic level. Hence, Eq.~\eqref{eq:FinalWeakForm} states equality of work at macro- and micro-scales.

The weak form in Eq.~\eqref{eq:FinalWeakForm} will be complemented with either Dirichlet or Neumann \emph{macroscopic} conditions on each of the components of the macroscopic energy-conjugate variables $\overline{\Strain}-\overline{\bm{\sigma}}$ and $\overline{\E}-\overline{\toVect{D}}$, as shown next. Therefore, in practice, the right hand side of the system of equations to be solved is a priori known.

The variations $\delta\Displacement(\x)$ and $\delta\phi(\x)$ in Eq.~\eqref{eq:FinalWeakForm} are high-order generalized-periodic functions with the same regularity as the primal unknowns $\Displacement(\x)$ and $\phi(\x)$, that is, their components are $\mathcal{C}^1$-continuous in $\Omega^{\textnormal{RVE}}$. Hence, they can be expressed in terms of two independent sets of variations as
\begin{subequations}
\label{eq:DeltaDecomposition}
    \begin{align}
        \delta\Displacement(\x)&=\delta\Displacement^\text{P}(\x) + \delta\overline{\Strain} \cdot \x, \\
        \delta\phi(\x)&=\delta\phi^\text{P}(\x)- \delta\overline{\E}\cdot \x,
    \end{align}
\end{subequations}
analogously to Eq.~\eqref{eq:StateDecomposition}, where the microscopic variations $\delta\Displacement^\text{P}(\x)$ and $\delta\phi^\text{P}(\x)$ are high-order periodic functions with $\mathcal{C}^1$-continuous components in $\Omega^{\textnormal{RVE}}$, and the macroscopic variations
$\delta\overline{\Strain}$ and $\delta\overline{\E}$ are a constant symmetric matrix and a constant vector, respectively. It is easy to verify that
\begin{subequations}
	\label{eq:DeltaDecomposition2}
	\begin{align}
		\overline{\Strain}\left(\delta\Displacement\right) &= \delta\overline{\Strain}, \\
		\overline{\E}(\delta\phi) &= \delta\overline{\E},
	\end{align}
\end{subequations}
that is, the variations in the right hand side of Eq.~\eqref{eq:FinalWeakForm} are purely macroscopic. By introducing Eqs.~\eqref{eq:DeltaDecomposition} and \eqref{eq:DeltaDecomposition2} into Eq.~\eqref{eq:FinalWeakForm}, a set of two Eqs.~is obtained as follows:
\begin{subequations}\label{eq:FinalWeakForm2}\begin{align}
	&\label{eq:FinalWeakForm2a}
	\frac{1}{|\Omega^{\textnormal{RVE}}|}\int_{\Omega}\left(\cauchyStress_{ij}\!\left(\Displacement,\phi\right)\strain_{ij}\!\left(\delta\Displacement^\text{P}\right)-\electricDisp_l\!\left(\Displacement,\phi\right) E_l\!\left(\delta\phi^\text{P}\right)+\hyperStress_{ijk}\!\left(\Displacement,\phi\right)\strain_{ij,k}\!\left(\delta\Displacement^\text{P}\right)-\widetilde{D}_{lm}\!\left(\Displacement,\phi\right)E_{l,m}\!\left(\delta\phi^\text{P}\right)\right)\!\dd \Omega
	=	0,
\\&\label{eq:FinalWeakForm2b}
	\frac{1}{|\Omega^{\textnormal{RVE}}|}\int_{\Omega}\left(\cauchyStress_{ij}\!\left(\Displacement,\phi\right)\delta\overline{\strain}_{ij}-\electricDisp_l\!\left(\Displacement,\phi\right) \delta\overline{E}_l \right)\!\dd \Omega
	=
	\overline{\sigma}_{ij}\!\left(\Displacement,\phi\right)\delta\overline{\strain}_{ij} - \overline{D}_l\!\left(\Displacement,\phi\right)\!\delta\overline{E}_l,
\end{align}\end{subequations}
for every admissible microscopic variations $\delta\Displacement^\text{P}(\x)$, $\delta\phi^\text{P}(\x)$, and macroscopic ones $\delta\overline{\Strain}$, $\delta\overline{\E}$, respectively.
Eq.~\eqref{eq:FinalWeakForm2a} states that purely microscopic variations of the state variables must not introduce work into the system, whereas Eq.~\eqref{eq:FinalWeakForm2b} reveals the mathematical relation between microscopic/macroscopic stress/electric fields as
\begin{subequations}\label{eq:FinalWeakForm3}\begin{align}
		&\label{eq:FinalWeakForm3a}
		\frac{1}{|\Omega^{\textnormal{RVE}}|}\int_{\Omega}\cauchyStress_{ij}\!\left(\Displacement,\phi\right)\!\dd \Omega
		=
		\overline{\sigma}_{ij}\!\left(\Displacement,\phi\right),
		\\&\label{eq:FinalWeakForm3b}
		\frac{1}{|\Omega^{\textnormal{RVE}}|}\int_{\Omega}\electricDisp_l\!\left(\Displacement,\phi\right)\!\dd \Omega
		=
		\overline{D}_l\!\left(\Displacement,\phi\right),
\end{align}\end{subequations}
i.e., macroscopic stress/electric fields are essentially the spatial average of the microscopic (local) ones. Note that Eq.~\eqref{eq:FinalWeakForm2b} is macroscopic in nature, hence it does not depend on $\x$, i.e.~the number of Eqs.~in \eqref{eq:FinalWeakForm2b} after spatial discretization will not depend on the mesh resolution.

Rewriting of the weak form, as in Eqs.~in \eqref{eq:FinalWeakForm2}, is very convenient to impose \emph{macroscopic conditions} on \emph{each component} of the macroscopic state variables. Dirichlet macroscopic conditions (e.g.~known values of $\overline{\strain}_{ij}$ or $\overline{E}_l$) can be strongly enforced, implying that the corresponding macroscopic variations (e.g.~$\delta\overline{\strain}_{ij}$ or $\delta\overline{E}_l$) vanish, and hence the number of Eqs.~in \eqref{eq:FinalWeakForm2b} after spatial discretization decreases accordingly. In turn, Neumann macroscopic conditions (e.g.~known values of $\overline{\sigma}_{ij}$ or $\overline{D}_l$) are trivially enforced by substituting the known value of the stress/electric field in the right hand side of Eq.~\eqref{eq:FinalWeakForm2b}. 

Even after imposing macroscopic conditions, the primal variables $\Displacement(\x)$, $\phi(\x)$ are well-defined up to rigid body translations, given that no \emph{microscopic} Dirichlet boundary conditions are enforced. In order to obtain a unique solution, it is sufficient imposing vanishing displacement and electric potential at an arbitrary point in $\Omega$.

\section{Numerical discretization and level-set geometry description}

Following the framework introduced by \cite{codony2019immersed}, we consider a representation of the displacement and potential fields with uniform Cartesian B-spline basis functions \citep{Rogers2001,Piegl2012,deBoor2001}, which are element-wise multivariate polynomial bases of degree $p$ with $\mathcal{C}^{p-1}$ continuity across elements. In order to fulfill the high-order generalized periodicity constraint in $\Omega$, we simply transform the uniform B-spline approximation space to a high-order generalized-periodic B-spline approximation space \citep{barceloinprep} by enforcing certain linear constraints on the basis functions intersecting the unit cell boundaries. 

\begin{figure}  
	\centering
	\begin{minipage}{0.66\textwidth}
	\subfigure[]
	{\includegraphics[width=0.5\textwidth,clip,keepaspectratio,angle=0]{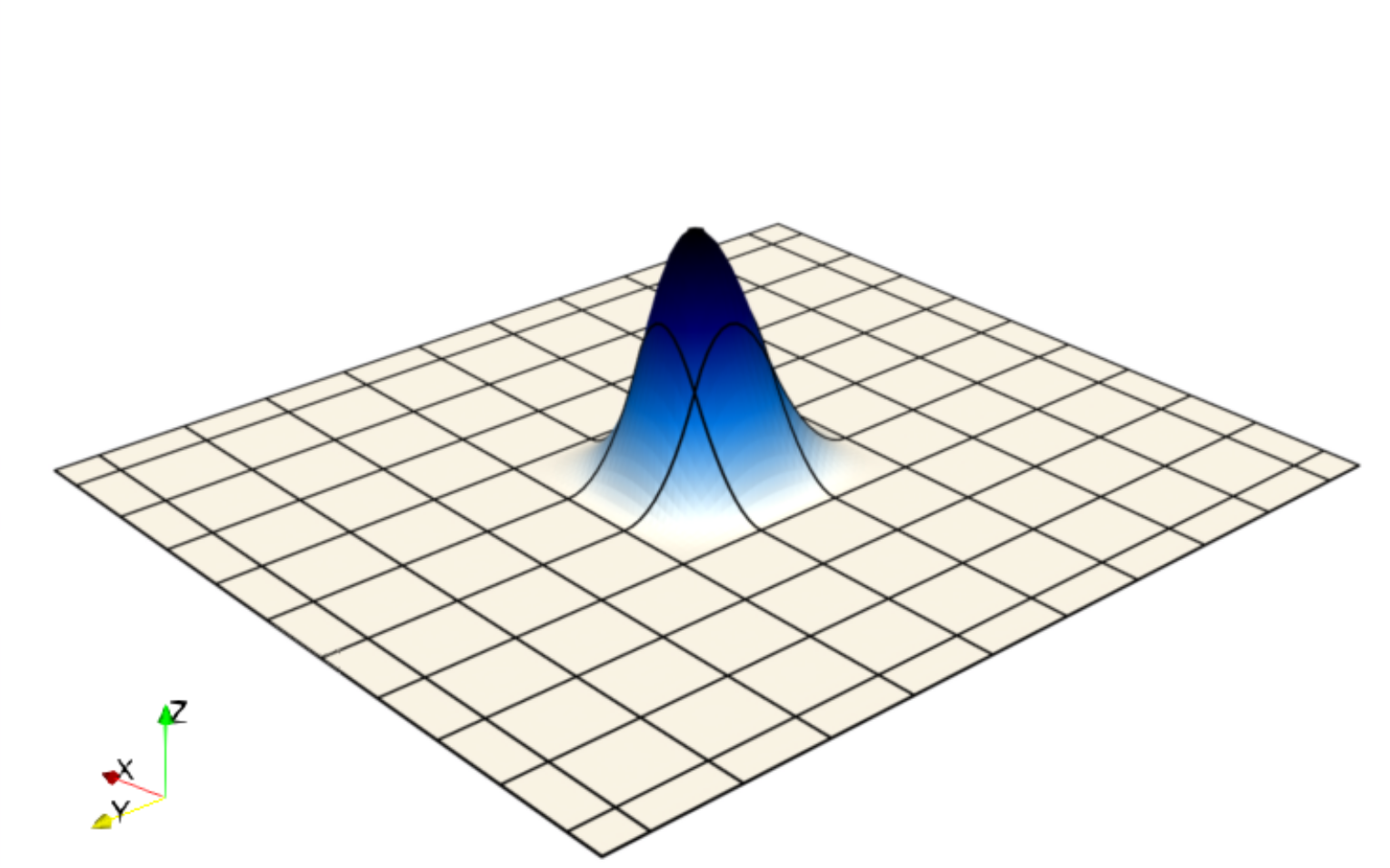}}
	\subfigure[]
	{\includegraphics[width=0.5\textwidth,clip,keepaspectratio,angle=0]{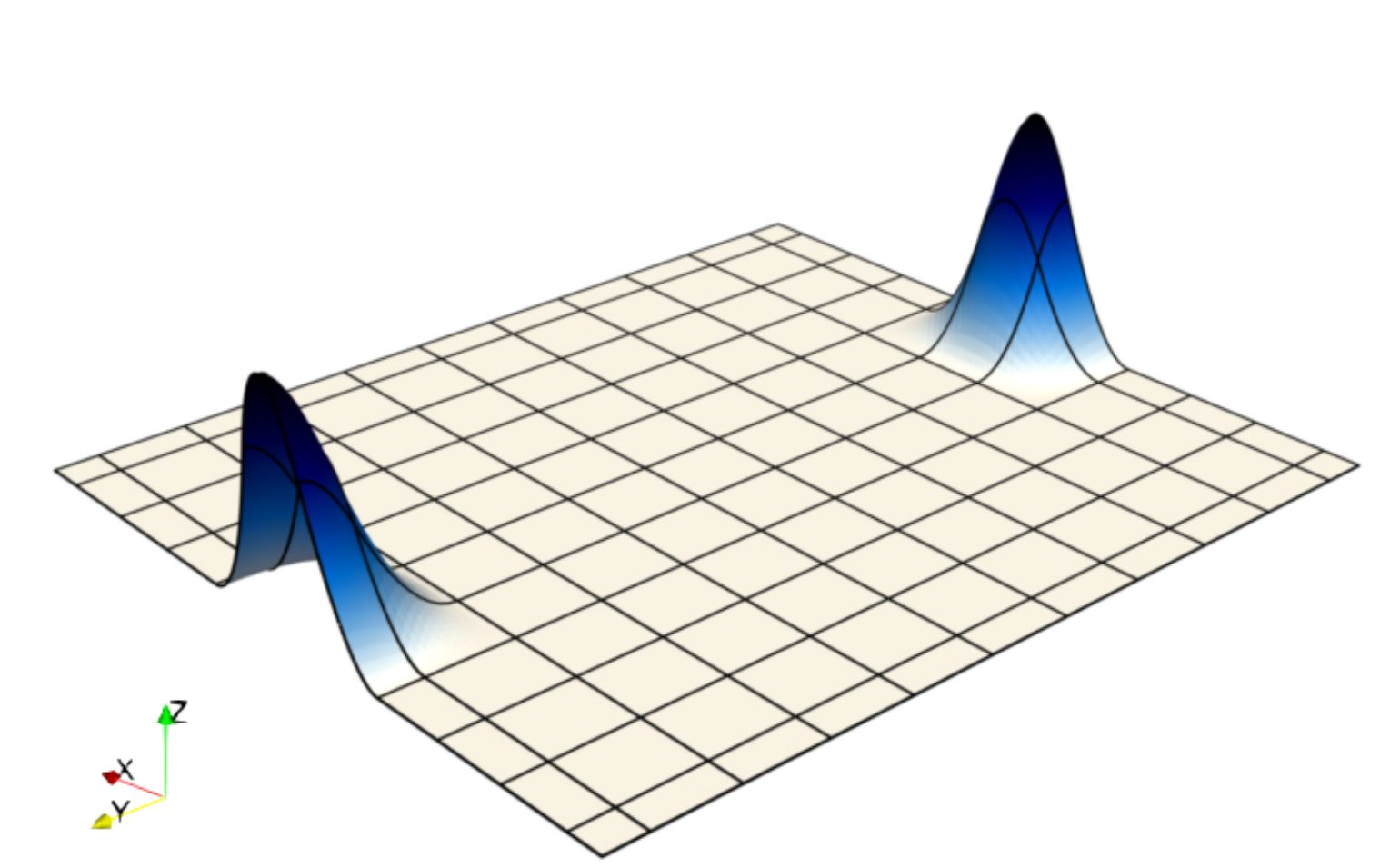}}\\
	\subfigure[]
	{\includegraphics[width=0.5\textwidth,clip,keepaspectratio,angle=0]{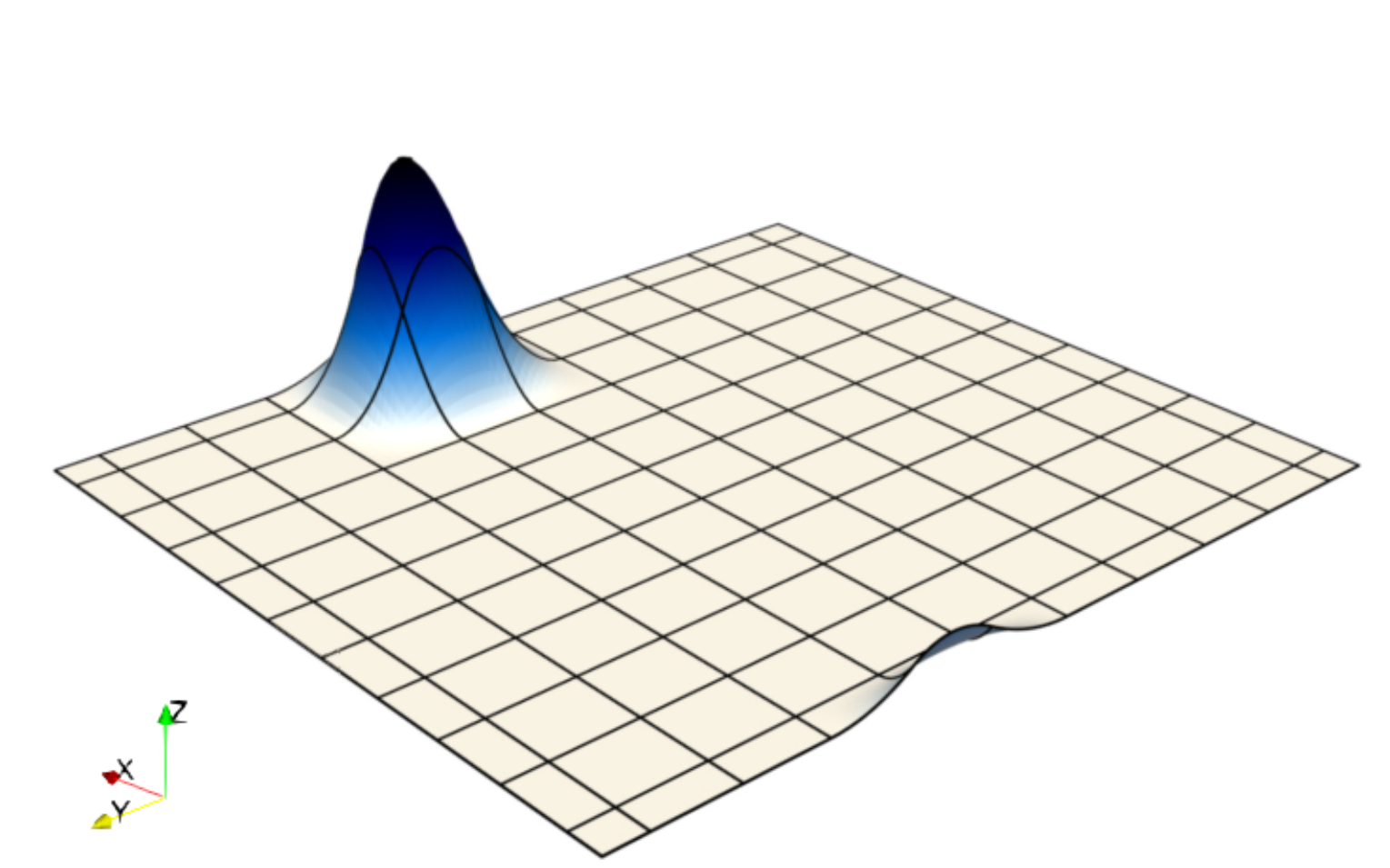}}
	\subfigure[]
	{\includegraphics[width=0.5\textwidth,clip,keepaspectratio,angle=0]{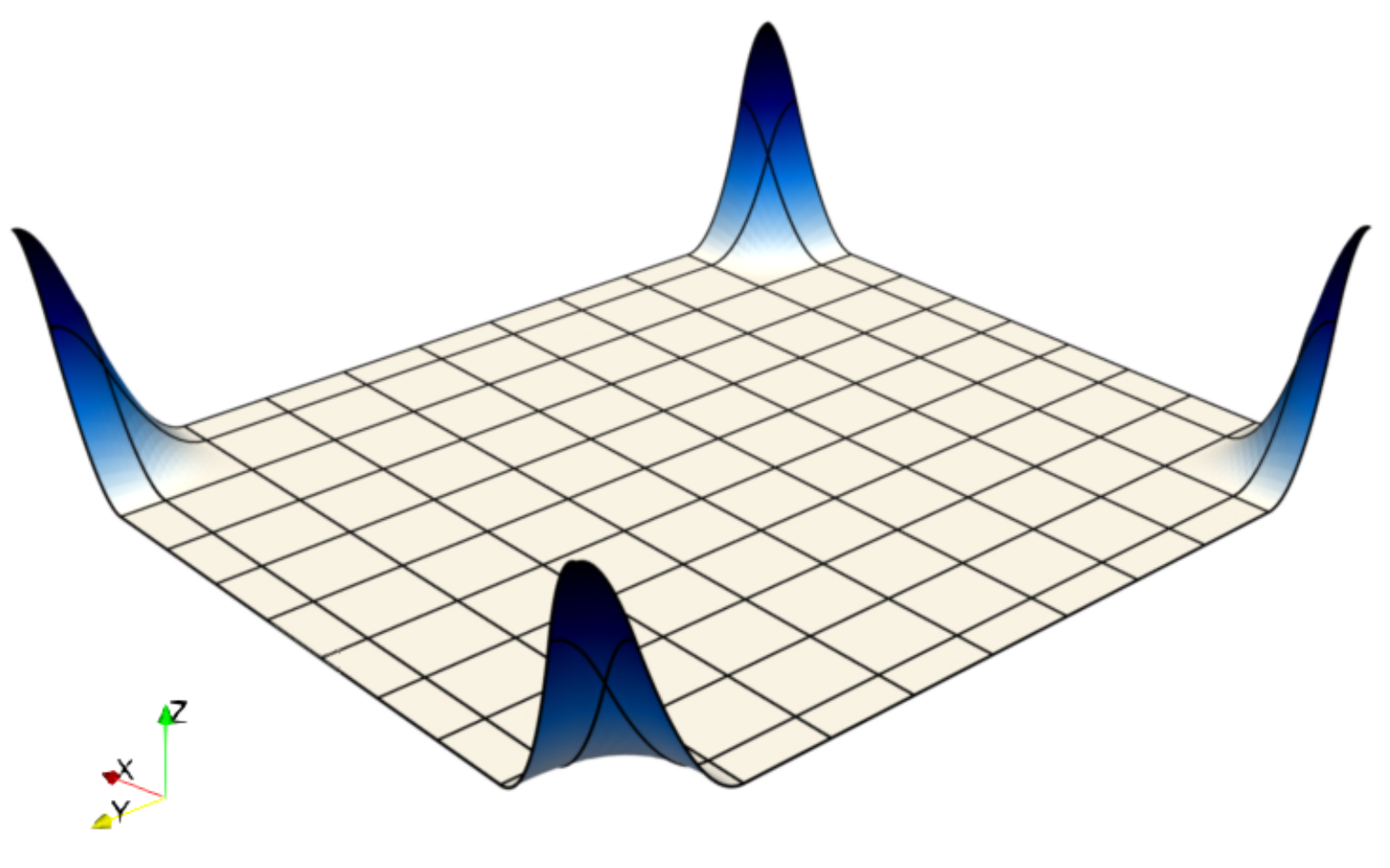}}
	\end{minipage}
	\begin{minipage}{0.33\textwidth}
	\subfigure[]
	{\includegraphics[width=\textwidth,clip,keepaspectratio,angle=0]{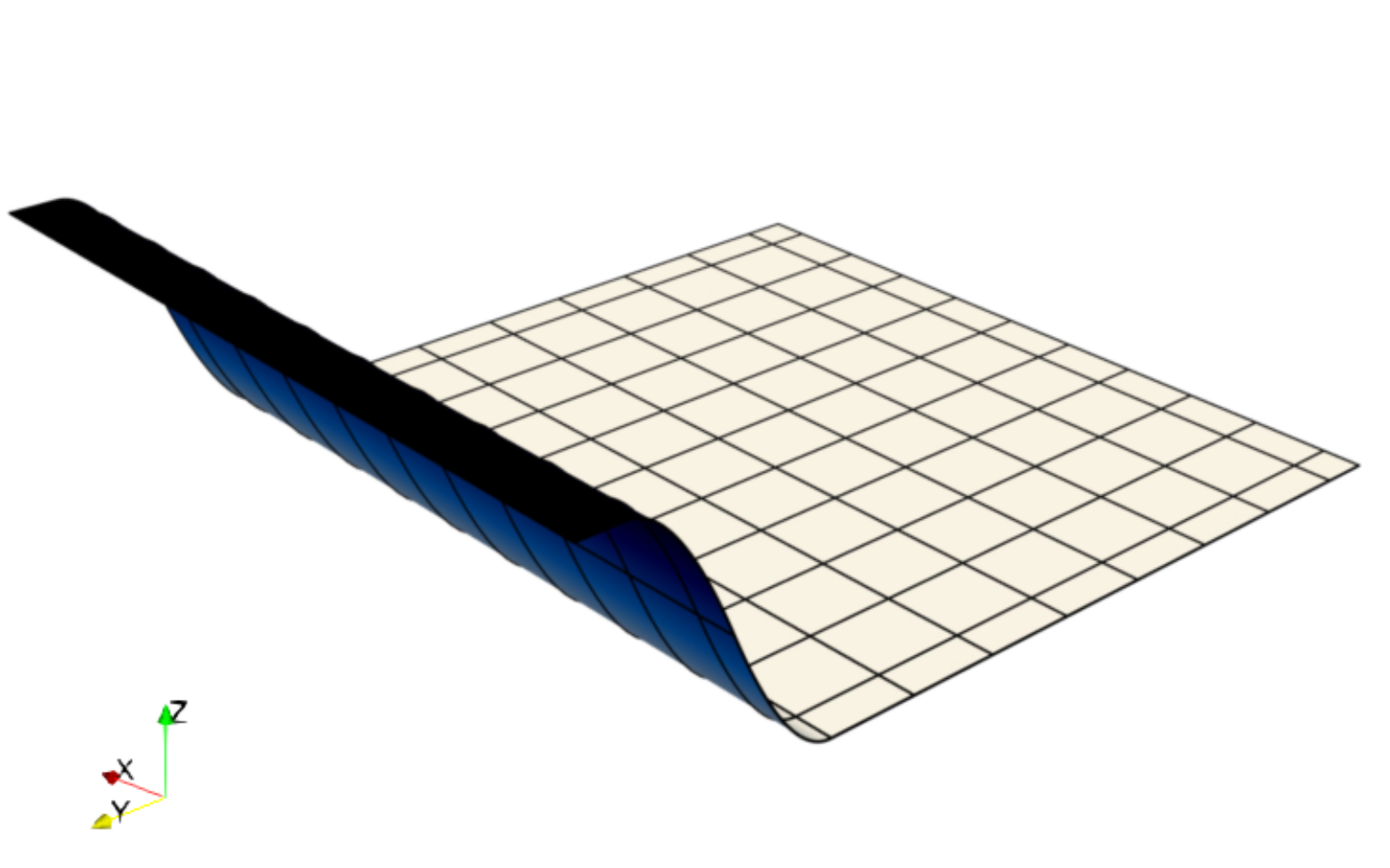}}
	\end{minipage}
	\caption{\label{fig:basis}Some of the basis functions $N_i(\x)$ of degree $p=2$ spanning the high-order generalized-periodic approximation space on the computational mesh. (a)--(d) are periodic multivariate B-spline functions, whereas (e) is a smooth univariate ramp function.} 	
\end{figure}

The resulting approximation space is a uniform Cartesian mesh with element sizes $(h_x, h_y )$, due to the tensor product nature of the basis functions. The element sizes $h_\xi$ are a subdivision of the unit cell sizes $L_\xi$, so that $L_\xi = n_\xi h_\xi$, where $n_\xi\in\mathbb{N}$ are the number of elements in each direction (taking periodicity into account)  and  $\xi = \{x,y\}$. The approximation space is spanned by two kind of basis functions, illustrated in Fig.~\ref{fig:basis}: i) periodic multivariate B-spline functions with support on $(p+1)\times(p+1)$ elements, which form a \emph{high-order periodic} space, and ii) smooth ramp functions with support on $(p+1)\times n_\xi$ elements and varying along one dimension only, which make the approximation space \emph{generalized}-periodic. Every basis function inherits the $\mathcal{C}^{p-1}$ continuity of the original uniform B-spline approximation space of degree $p$.

Mathematically, the unknowns of the problem $\Displacement(\x)$ and $\phi(\x)$ are approximated as 
\begin{subequations}
    \begin{align}
        [\Displacement(\bm{x})]_\xi\approx [\Displacement^h(\bm{x})]_\xi&=N_i(\bm{x})a^u_{i\xi},\\
\phi(\bm{x})\approx\phi^h(\bm{x})&=N_i(\bm{x})a^\phi_{i},
    \end{align}
\end{subequations}
where $N_i(\x)$ are the basis functions in physical space and $\{\bm{a}^u,\bm{a}^\phi\}$ are the degrees of freedom of $\Displacement^h$ and $\phi^h$.

The mesh is generally unfitted to $\Omega$, which can have an arbitrary shape. Some mesh elements are arbitrarily cut by $\partial\Omega$, and an appropriate strategy has to be implemented for their numerical integration and stabilization. In this work, we adopt an implicit description of the geometry $\Omega$ by means of a level-set function, combined with a non-conforming quadtree numerical integration scheme \citep{rank2012geometric}. This approach leads to a simple implementation and a relatively low computational cost. In order to alleviate severe ill-conditioning of the resulting algebraic system of equations due to cut cells, we consider the extended B-spline algorithm \citep{hollig2001weighted}, which consists on modifying the approximation space near $\partial\Omega$ by binding basis functions whose support is mostly outside of $\Omega$ with adjacent basis functions. See Refs.~\cite{codony2019immersed,codony2021mathematical} for further details.
 
The level-set function $\varphi(\x)$ is constructed from a linear combination of certain basis functions $M(\x)$ and coefficients $\bm{\varphi}$, and the zero level of $\varphi(\x)$ describes the physical boundary of the domain $\Omega$. Mathematically,
    \begin{align}
        \Omega = \{\x\in\Omega^\textrm{RVE}|\varphi(\x)>0\},
        \qquad
        \text{ where }
        \qquad
        \varphi(\x)=M_i(\bm{x})\bm{\varphi}_{i}.
    \end{align}
The level set function $\varphi(\x)$ inherits the periodicity requirements on $\Omega$. Therefore, the basis functions $M(\x)$ must be periodic as well. In the present context, the simplest choice is to consider high-order multivariate periodic B-spline basis, which will generate a smooth boundary of $\Omega$. Note that $M(\x)$ span a periodic space (geometry space), whereas $N(\x)$ span a generalized-periodic one (computational space). Both sets of basis functions are independent and, in particular, it is of practical interest them having different resolutions, i.e. different ``element'' sizes. We typically consider a finer computational space than the geometry space, which has element sizes $\ell_\xi\geq h_\xi$. Specifically, in order to exploit the advantages of the presented numerical scheme, we restrict ourselves to a computational mesh constructed as a subdivision of the geometry mesh, in such a way that $(\ell_x,\ell_y) = (s_xh_x,s_yh_y)$, where $s_\xi\in\mathbb{N}$ are the number of subdivisions in each direction. Note that $\ell_\xi$ have to be also a subdivision of the unit cell sizes, so that the number of elements per dimension in the geometry mesh is $n_\xi/s_\xi\in\mathbb{N}$. Having a nested hierarchy of computational and geometry meshes allows us to consider just $s_x\times s_y$ reference elements for the evaluation of the level set function at the integration points of the quad-tree scheme in the computational mesh elements. This enables a very efficient strategy to classify elements in the computational mesh depending whether they are within $\Omega$, outside $\Omega$, or cut by $\partial\Omega$, a step that is required in any unfitted discretization method, as well as an efficient evaluation of the weak form at the integration points.

For a fixed geometry $\Omega$ given by a set of coefficients $\bm{\varphi}$, the weak form described in Section \ref{sec:2} is discretized on the computational mesh, by assembling the different contributions from interior and cut elements. The macroscopic Dirichlet conditions are enforced strongly by prescribing the degrees of freedom that multiply the corresponding ramp basis functions {and the macroscopic Neumann conditions are directly applied at the right hand side}. As a result, an algebraic linear system of equations is obtained as
\begin{equation}
    \mathbf{K}(\bm{\varphi})\cdot\mathbf{a}(\bm{\varphi}) = \mathbf{f}(\bm{\varphi}),
\end{equation}
which one can solve to obtain {$\mathbf{a}(\bm{\varphi})=\{\bm{a}^u(\bm{\varphi}),\bm{a}^\phi(\bm{\varphi})\}$}.

\subsection{Identifying valid level set configurations}
\begin{figure}  
	\centering
	\subfigure[]
	{\includegraphics[width=0.30\textwidth,clip,keepaspectratio,angle=0]{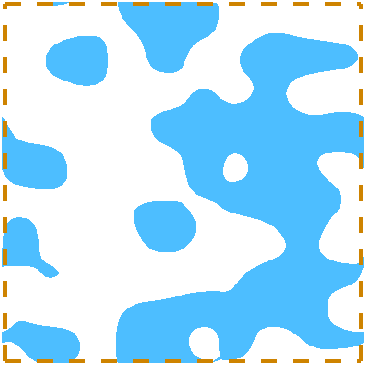}}\;\;\;
	\subfigure[]
	{\includegraphics[width=0.30\textwidth,clip,keepaspectratio,angle=0]{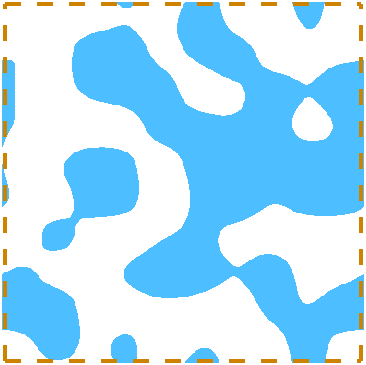}}	\;\;\;
	\subfigure[]
	{\includegraphics[width=0.30\textwidth,clip,keepaspectratio,angle=0]{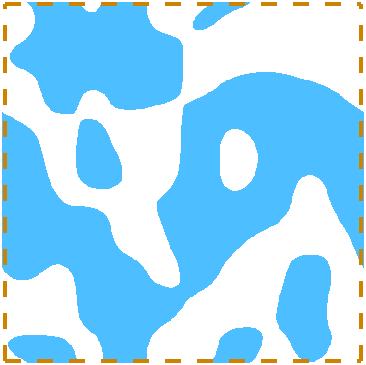}}	
	\caption{Three random domains generated with a 10x10 quadratic level-set.} 
	\label{fig:randls}
\end{figure}

\begin{figure}  
	\centering
	\subfigure[]
	{\includegraphics[width=0.4\textwidth,clip,keepaspectratio,angle=0]{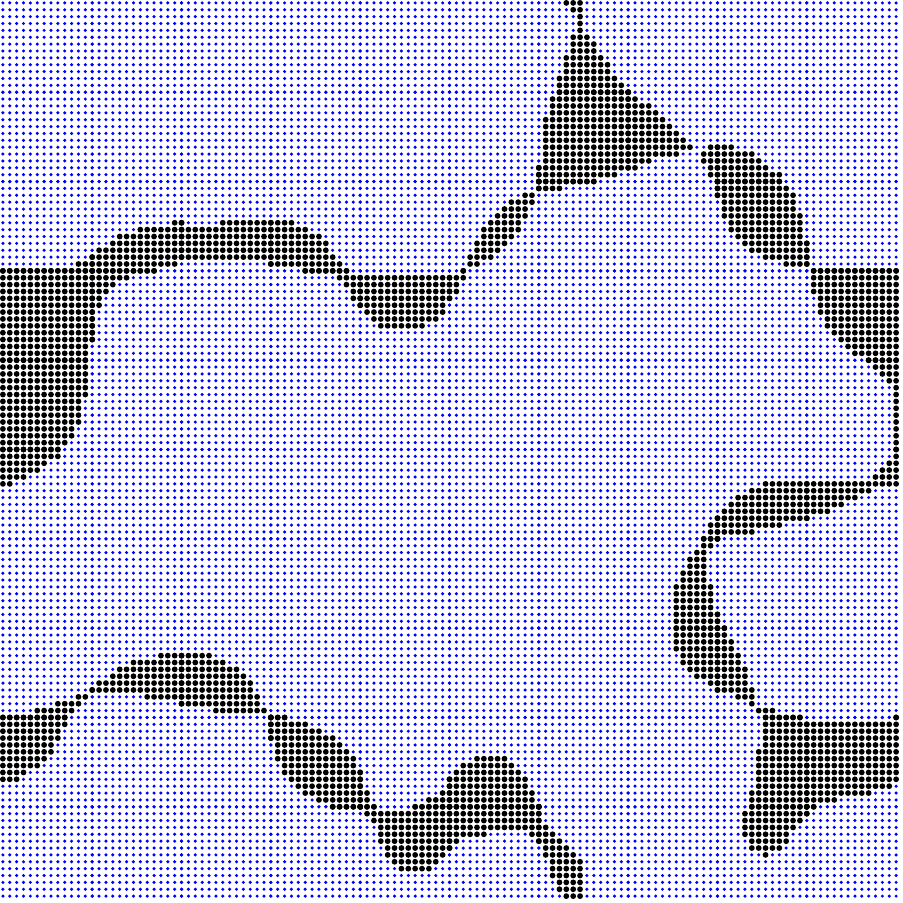}}
	\subfigure[]
	{\includegraphics[width=0.4\textwidth,clip,keepaspectratio,angle=0]{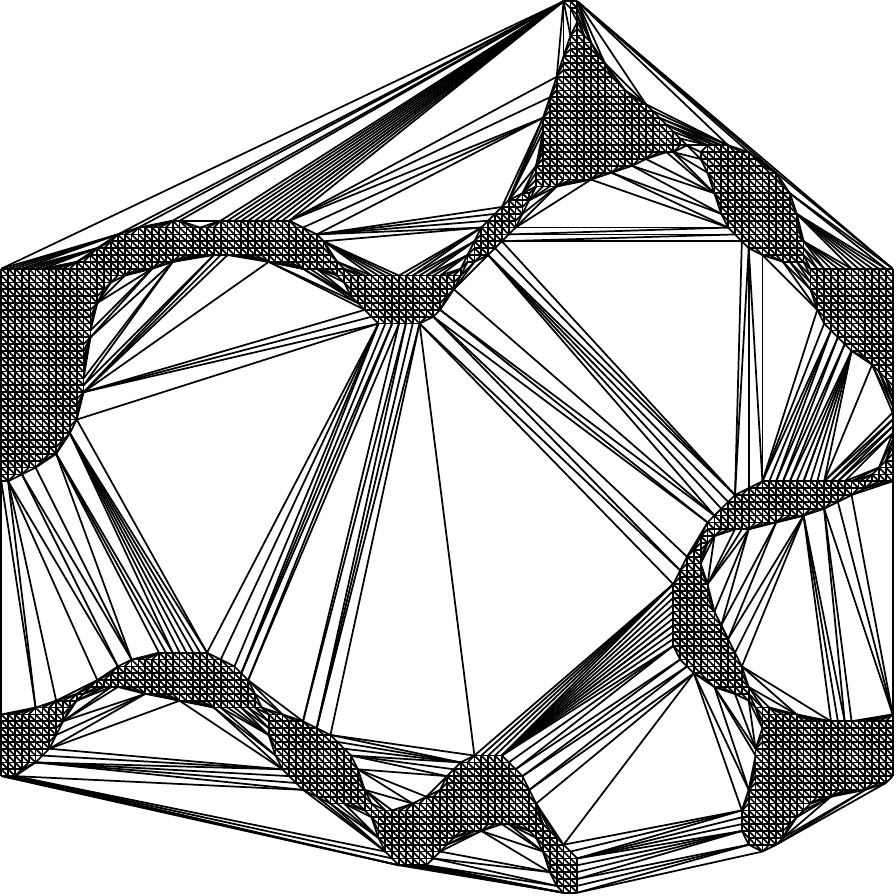}}	
	\subfigure[]
	{\includegraphics[width=0.4\textwidth,clip,keepaspectratio,angle=0]{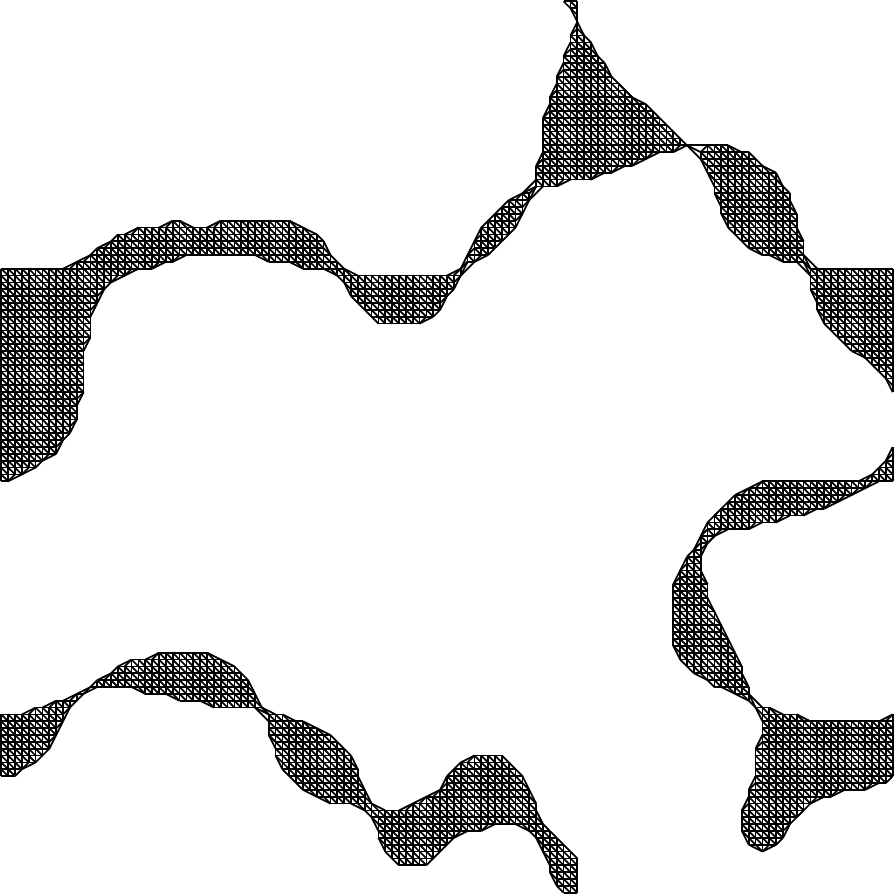}}	
	\subfigure[]
	{\includegraphics[width=0.4\textwidth,clip,keepaspectratio,angle=0]{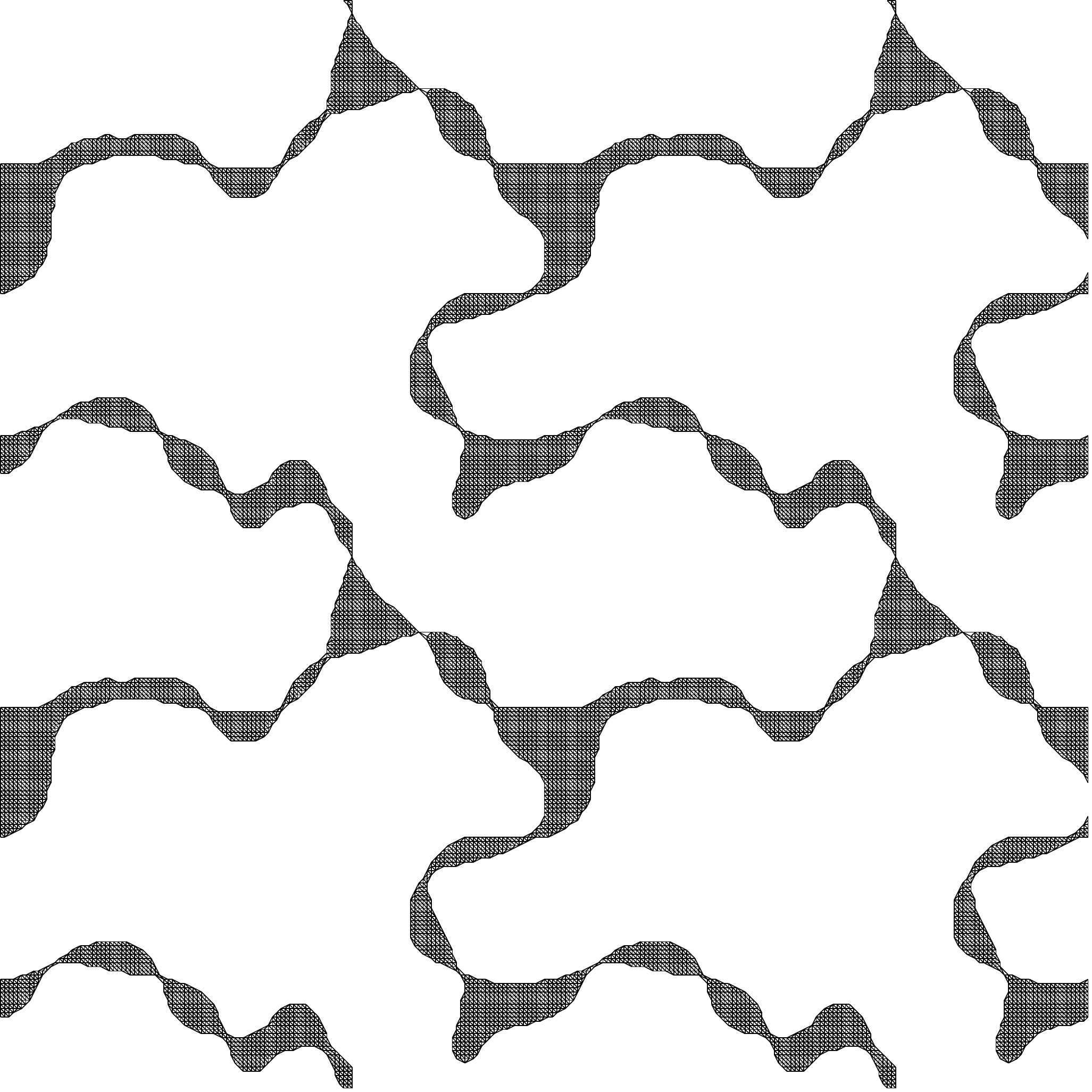}}	
	\caption{Procedure to determine topologically valid geometries. The domain is first sampled with a cloud of points and only those within the level set   $\varphi(\x)$ (in black) are maintained (a); then the Delaunay triangulation is computed (b); the alpha-shape is extracted (c); and the domain is duplicated to count the number of disconnected subdomains (d). We note that for the purpose of this illustration, a low number of sampling points and a slightly higher value of $\alpha$ have been used.} 
	\label{fig:sampling}
\end{figure}

As we can observe in Fig.~\ref{fig:randls}, a randomly generated level set description of $\Omega$ {usually} yields domains with disconnected parts{, which are not} valid topologies for the architected materials {and would lead to ill-posed problems with arbitrary generalized-periodic solutions}.
To {avoid non-valid geometries}, the following algorithm is used to identify the number of connected parts or groups $N_g$ present in the structure: 
\begin{itemize}
	\item the domain described by the level set is first sampled with a point cloud by considering a fixed tensor product grid in each level set element (Fig.~\ref{fig:sampling}a); that is, the level set function is evaluated on a regular grid and the points with $\varphi(\x) \geq0$ are maintained;
	\item 
	the Delaunay triangulation is then constructed on this point cloud (Fig.~\ref{fig:sampling}b) and a shape containing the convex-hull of the cloud is obtained;
	\item the alpha shape criterion \citep{akkiraju1995alpha} is used to extract an approximation of the shape of the domain. The idea of this method is to remove from the triangulation all the triangles whose circumradius is bigger than a fixed $\alpha$ value, which in this case corresponds to the resolution used for the level set sampling;
	\item starting from the alpha-shape triangulation of the domain (Fig.~\ref{fig:sampling}c), the number of groups $N_g$ is counted{, as explained next for an implementation in \textsc{Matlab}\textsuperscript{\circledR}}.
\end{itemize}

For this last step, it is convenient to convert the triangulation into a graph object in \textsc{Matlab}\textsuperscript{\circledR}, \emph{account for its periodicity by modifying its connectivity}, and use the \texttt{conncomp} function \citep{tarjan1972depth}. To account for the periodicity of the RVE, the connectivity matrix of the triangulation is modified by identifying the vertices of the triangulation lying on the RVE boundaries and their periodic images with the same index.
As an example, a naive count of the number of groups in Fig.~\ref{fig:sampling}d without taking periodicity into account yields $N_g=4$, indicating that the structure is disconnected. Instead, a count of the  number of groups with a periodicity-aware connectivity matrix gives the correct number $N_g=1$ that indicates a globally connected structure.

\begin{figure}  
	\centering
	\subfigure[Island of material ($1\times1$)]
	{\hspace*{0.050\textwidth}\includegraphics[width=0.12\textwidth,clip,keepaspectratio,angle=0]{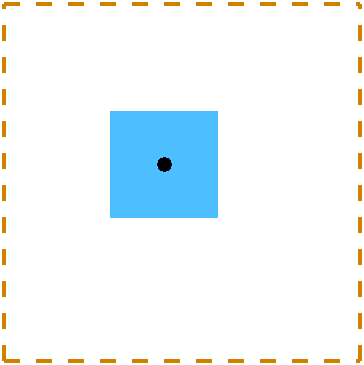}\hspace*{0.050\textwidth}}\;\;\;
	\subfigure[Horizontal domain ($1\times1$)]
	{\hspace*{0.050\textwidth}\includegraphics[width=0.12\textwidth,clip,keepaspectratio,angle=0]{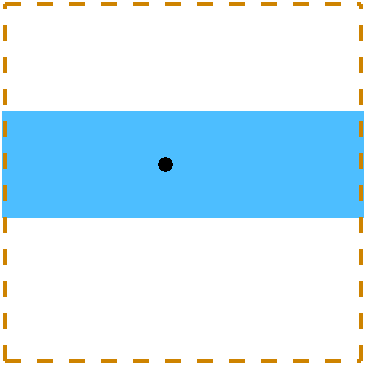}\hspace*{0.050\textwidth}}\;\;\;
	\subfigure[Vertical domain ($1\times1$)]
	{\hspace*{0.050\textwidth}\includegraphics[width=0.12\textwidth,clip,keepaspectratio,angle=0]{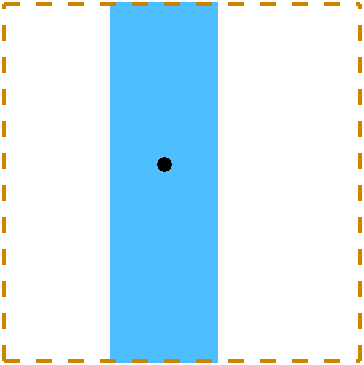}\hspace*{0.050\textwidth}}\;\;\;
		\subfigure[Valid configuration ($1\times1$)]
	{\hspace*{0.050\textwidth}\includegraphics[width=0.12\textwidth,clip,keepaspectratio,angle=0]{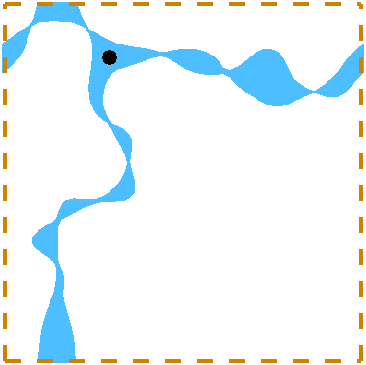}\hspace*{0.050\textwidth}}
		\subfigure[Island of material ($2\times2$)]
	{\includegraphics[width=0.22\textwidth,clip,keepaspectratio,angle=0]{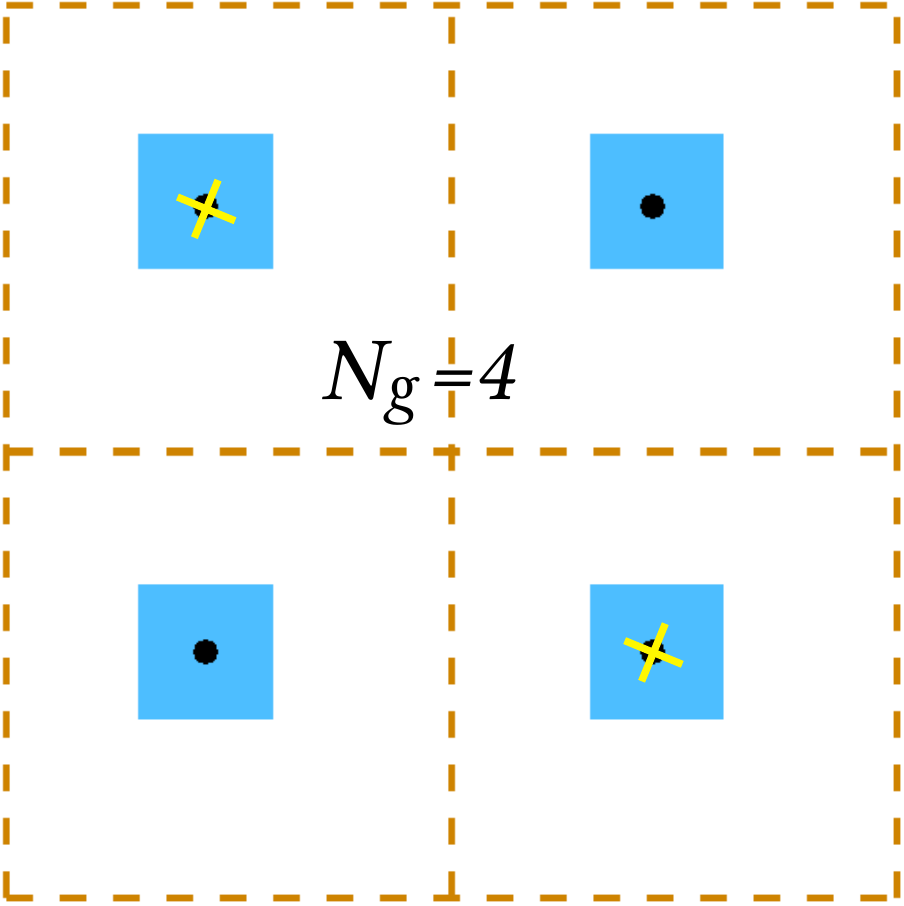}}\;\;\;
		\subfigure[Horizontal domain ($2\times2$)]
	{\includegraphics[width=0.22\textwidth,clip,keepaspectratio,angle=0]{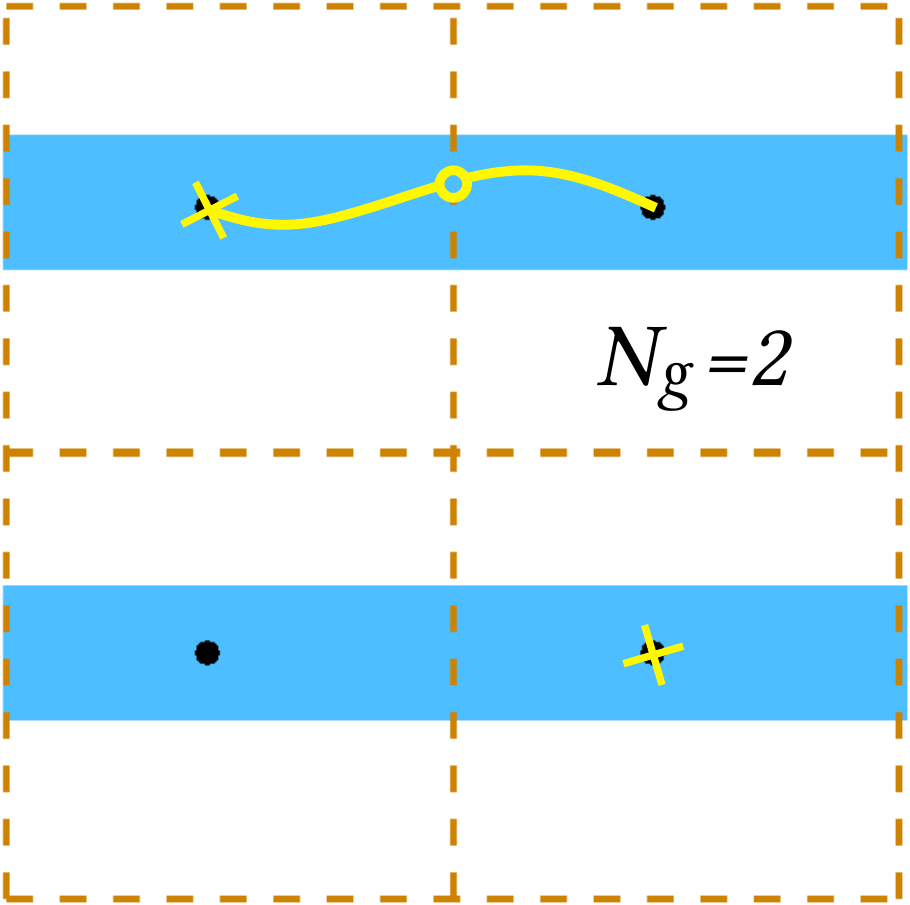}}\;\;\;
		\subfigure[Vertical domain ($2\times2$)]
	{\includegraphics[width=0.22\textwidth,clip,keepaspectratio,angle=0]{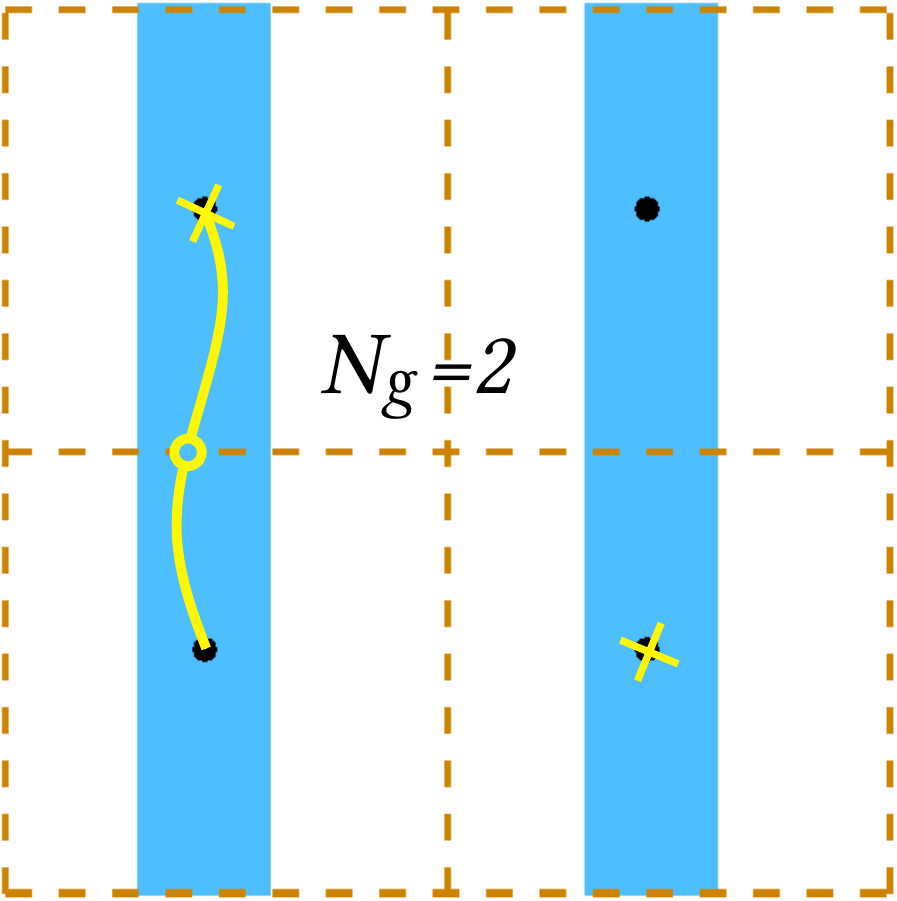}}\;\;\;
		\subfigure[Valid configuration ($2\times2$)]
	{\includegraphics[width=0.22\textwidth,clip,keepaspectratio,angle=0]{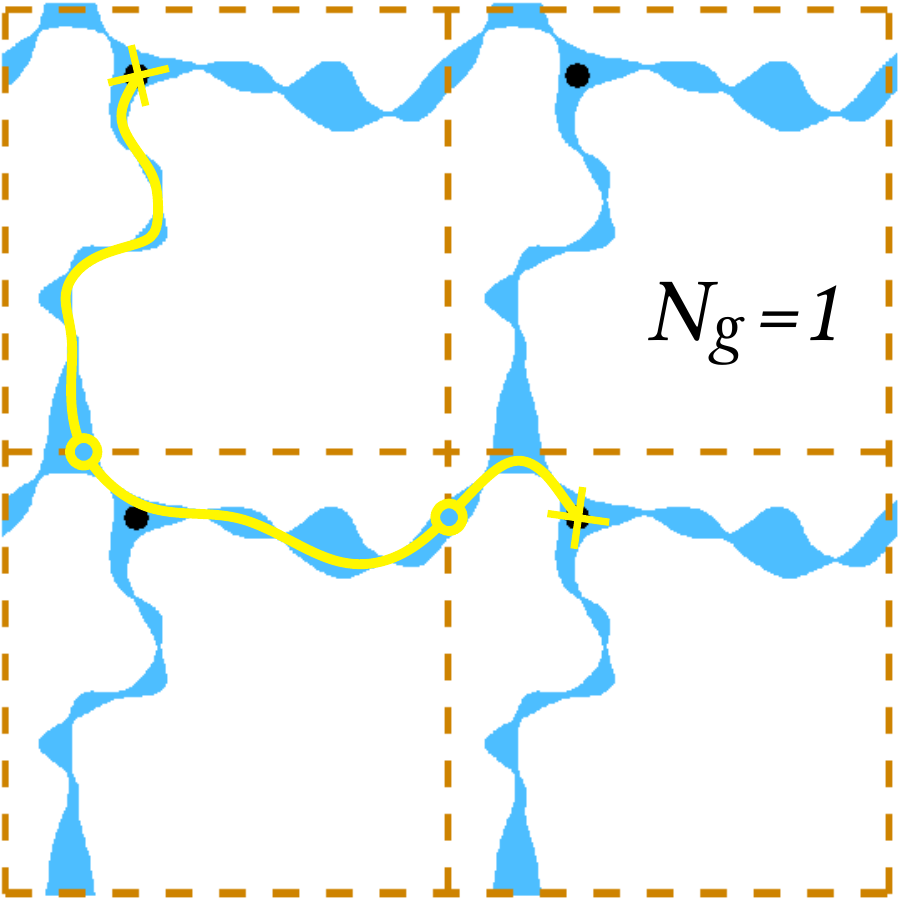}}	
	\caption{Four different configurations in a single ($1\times1$) and replicated ($2\times2$) RVE. While $N_g=1$ in the single RVE for all cases, $N_g$ in the replicated RVE might differ. Only in the valid configuration of (d) and (h) it is possible to create a path (yellow line) between each point in the domain and an image of itself in the opposite RVE (yellow crosses), while intersecting every RVE boundary (yellow circles).	} 
	\label{fig:isola}
\end{figure}

However, having a number of groups $N_g=1$ in the RVE is not sufficient to assert that the periodic structure is actually connected.
On top of that, it is also required that, for each point in the RVE, it is possible to create a path that connects this point with itself while intersecting every boundary of the RVE.
A straightforward way to check whether this condition is fulfilled is {duplicating} the RVE in both the $x$ and $y$ directions, and just {counting} the groups therein. Fig.~\ref{fig:isola} illustrates this {procedure} in three disconnected structures 
(i.e.~an isolated group of material, horizontal or vertical domains in Figs.~\ref{fig:isola}a-c), and a connected one (Fig.~\ref{fig:isola}d).

\section{Optimization with genetic algorithms}

Following classical piezoelectricity \citep{ikeda1996fundamentals} and the approach used by \cite{mocci2021geometrically}, we define four apparent piezoelectric coupling coefficients of
our metamaterials depending on the electromechanical ensemble:
\begin{equation} \label{coeff}
	\overline{h}=-\dfrac{\partial\overline{E}_y }{\partial\overline{\strain}_{yy}} \bigg|_{\overline{D}_y=0},\;\;\;
		\overline{d}=\dfrac{\partial\overline{\strain}_{yy} }{\partial\overline{E}_y} \bigg|_{\overline{\sigma}_{yy}=0},\;\;\;
			\overline{g}=-\dfrac{\partial\overline{E}_y }{\partial\overline{\sigma}_{yy}} \bigg|_{\overline{D}_y=0},\;\;\;
				\overline{e}=-\dfrac{\partial\overline{\sigma}_{yy}}{\partial\overline{E}_y}\bigg|_{\overline{\strain}_{yy}=0},
\end{equation}
being $\overline{h}$ and $\overline{g}$ pertinent to a sensing scenario {(mechanical input -- electric output)} and $\overline{d}$ and $\overline{e}$ to an actuation one {(electric input -- mechanical output). The complete sets of macroscopic conditions for each optimization problem are stated in Table \ref{tab:macrocond}.}

If we refer to one of the aforementioned coefficients as $\overline{c}$, the {considered} optimization problem {is}:
\begin{equation}\label{optim}
	\begin{aligned}
		\min_{   \bm{\varphi }    } \quad & -|\overline{c} (\bm{\varphi })|,  \\
		\textrm{subject to:}\quad &A_f(\bm{\varphi })  \leq A_\text{max}, \\
	&N_g(\bm{\varphi })=1,
	\end{aligned}
\end{equation}
where $\bm{\varphi } $ is the vector of control {variables} of the level set function describing the domain, $A_f(\bm{\varphi }) = |\Omega(\bm{\varphi })|/|\Omega^{\textnormal{RVE}}|$ is the area fraction of the structure, which is normally set to be smaller than a given fraction  $A_\text{max}$, and $N_g(\bm{\varphi })$ is the number of groups defined in the previous section.
The absolute value $|\overline{c} (\bm{\varphi })|$ is introduced to account for those configurations that, according to the definition given by Eq.~\eqref{coeff}, would have a negative coefficient but whose shape is still valid for the optimization purpose.

The aforementioned optimization problem can be solved using different  algorithms, each of them with advantages and drawbacks, as reviewed in the introduction. Here, we resort to genetic algorithms. This population-based approach examines a diverse set of solutions during the optimization process, a feature that is particularly advantageous in the context of meta-material design because it is less likely to converge prematurely to sub-optimal local solutions,  and hence more likely to led to unconventional designs. While the computational cost is higher with respect to other methodologies such as gradient based optimization, genetic algorithms are more easily implemented and parallelized.

As far as the numerical implementation is concerned, we use the \textsc{Matlab}\textsuperscript{\circledR} genetic algorithm toolbox \citep{chipperfield1995matlab}. However, we note that in feeding the optimization problem stated in Eq.~\eqref{optim} directly to the solver, the optimization procedure fails in most of the cases, since the initial randomly generated population in general does not verify both constraints, and then the algorithm converges to infeasible local minima either if a penalty or an augmented Lagrangian formulation is used.
We therefore resort to a more effective strategy, which consists in solving a multi-objective problem, where both the apparent piezoelectric coefficient and the area fraction are optimized:
\begin{equation}\label{optimpa}
	\begin{aligned}
		\min_{   \bm{\varphi }    } \quad & -|\overline{c} (\bm{\varphi })|, A_f(\bm{\varphi }),  \\
		\textrm{subject to:}\quad	& N_g(\bm{\varphi })=1.
	\end{aligned}
\end{equation}
This approach is not only more efficient from a numerical point of view but also more useful to understand the physics of the problem and how the different coefficients depend on the area fraction.
In fact, previous works consider the optimization of flexoelectric properties based on specific constrains such as area \cite{ghasemi2017level} or stress  \cite{ortigosa2022computational} but a multi-objective approach with focus on the Pareto fronts has not been presented yet to the best of the authors' knowledge and is the topics of the present work, with a detailed discussion following in the next section.

Also in the case of the Pareto formulation stated by Eq.~\eqref{optimpa}, we note that if a randomly generated initial population is used the algorithm often converges to a set of individuals located on a small fraction of the whole Pareto front.
Therefore, the initial population is enriched with a set of individuals computed by solving the modified Pareto problem:
\begin{equation} \label{optimbe}
	\begin{aligned}
		\min_{   \bm{\varphi }    } \quad & -|\overline{c} (\bm{\varphi })| +\beta A_f(\bm{\varphi }) + \gamma \left(N_g(\bm{\varphi })-1\right)^2,
	\end{aligned}
\end{equation}
where the Pareto parameter $\beta$ takes different values in a range that allows us to span the entire Pareto region. 
In this case, the constraint $N_g=1$ is imposed weakly using a penalty approach with a very high coefficient $\gamma$, such that even if the optimization starts with infeasible individuals, it progressively prefers those with smaller values of $N_g$ and gradually converges to valid configurations. In particular, a value of $\gamma=10^{20}$ is used.

\section{Results}

\subsection{Numerical identification of Pareto fronts}

The multi-objective problem stated in Eq.~\eqref{optimpa} is solved for each of the four apparent piezoelectric coefficients introduced in Eq.~\eqref{coeff}. We consider a square unit cell of size $4 \upmu$m. The material parameters are summarized in Table~\ref{tabmat} and correspond to those of a barium strontium titanate (BST) ferroelectric ceramic.
The complete sets of macroscopic conditions are stated in Table \ref{tab:macrocond} for each optimization problem.

\begin{table}[]
	\centering
		\caption{\label{tabmat}Material parameters used in the multi-objective optimization analyses.}
\begin{tabular}[t]{ |c| c| c| c|c|c|c|c|  }
	\hline
 	$Y$  & $\nu$ & $\ell_\textrm{mech}$ &$\ell_\textrm{elec}$  & $\epsilon$  & $\mu_\textrm{L}$ & $\mu_\textrm{T}$ &$\mu_\textrm{S}$ \\  
	 \; [Gpa] &   & [nm]& [nm] &  [nC/Vm] &  [$\upmu$C/m] & [$\upmu$C/m] & [$\upmu$C/m] \\ \hline
 	152 & 0.33 & 50 & 300 & 8 & 1.21 & 1.10 & 0.055\\ \hline
\end{tabular}
\end{table}

\begin{table}[]\centering
	\caption{\label{tab:macrocond}Macroscopic conditions for the multi-objective optimization analyses. A value 0 means that this corresponding quantity is constrained to 0, whereas (-) means that the corresponding quantity is a Lagrange multiplier for the constraint that does not participate in the relevant output.}
	\begin{tabular}{c|c|c|c|c|c|c|c|c|c|c|}
	\cline{2-11}
										 & $\overline{\strain}_{yy}$ & $\overline{\sigma}_{yy}$ & $\overline{E}_y$ & $\overline{D}_y$ & $\overline{\strain}_{xx}$ & $\overline{\sigma}_{xx}$ & $\overline{\strain}_{xy}$ & $\overline{\sigma}_{xy}$ & $\overline{E}_x$ & $\overline{D}_x$ \\ \hline
	\multicolumn{1}{|c|}{$\overline{h}$} & Input                     & -                         & Output           & 0                & 0                         & -                         & 0                         & -                         & 0                & -                \\
	\multicolumn{1}{|c|}{$\overline{d}$} & Output                    & 0                         & Input            & -                & 0                         & -                         & 0                         & -                         & 0                & -                \\
	\multicolumn{1}{|c|}{$\overline{g}$} & -                         & Input                     & Output           & 0                & 0                         & -                         & 0                         & -                         & 0                & -                \\
	\multicolumn{1}{|c|}{$\overline{e}$} & 0                         & Output                    & Input            & -                & 0                         & -                         & 0                         & -                         & 0                & -                \\ \hline
	\end{tabular}
	\end{table}

In the approach presented here, it is convenient to use different B-splines bases for the level set description of the geometry and for the  discretization of the flexoelectricity problem.
In genetic algorithms, a larger number of optimization variables requires in general a larger initial population and more generations. Therefore, a very fine discretization for the level set describing the geometry  quickly leads to computationally intractable problems. Nevertheless, a finer grid to solve flexoelectricity leads to better accuracy and therefore a better approximation of the objective function (the apparent piezoelectric coefficient) but does not increase so steeply the numerical cost of the optimization algorithm. Based on these considerations and on a set of preliminary numerical tests,
we found a good compromise by representing the geometry with a $10\times10$ quadratic ($p=2$) B-spline grid, whose control points are the variables to be optimized, and using a $100\times100$ cubic ($p=3$) B-spline grid for the numerical approximation.
The multi-objective optimization problems are solved considering a population of 2000 individuals and 2000 generations. 
 For all the cases, an initial randomly generated population is enriched with 30 seed individuals computed by solving the problem stated in Eq.~\eqref{optimbe}. 
In GA, new individuals are created at each generation through  a combination of cross-over and mutation of the elite individuals. In the specific problem solved here,  preliminary tests showed faster convergence with low cross-over fractions. This can be justified by the fact that, when combining two connected structures ($N_g=1$) via cross-over,  there is a higher chance of obtaining an invalid individual as compared to the mutation of a valid structure. Therefore a cross-over fraction of 0.2 is used to generate the initial seeds and the Pareto fronts.

\begin{figure}  
	\centering
\includegraphics[width=0.9\textwidth]{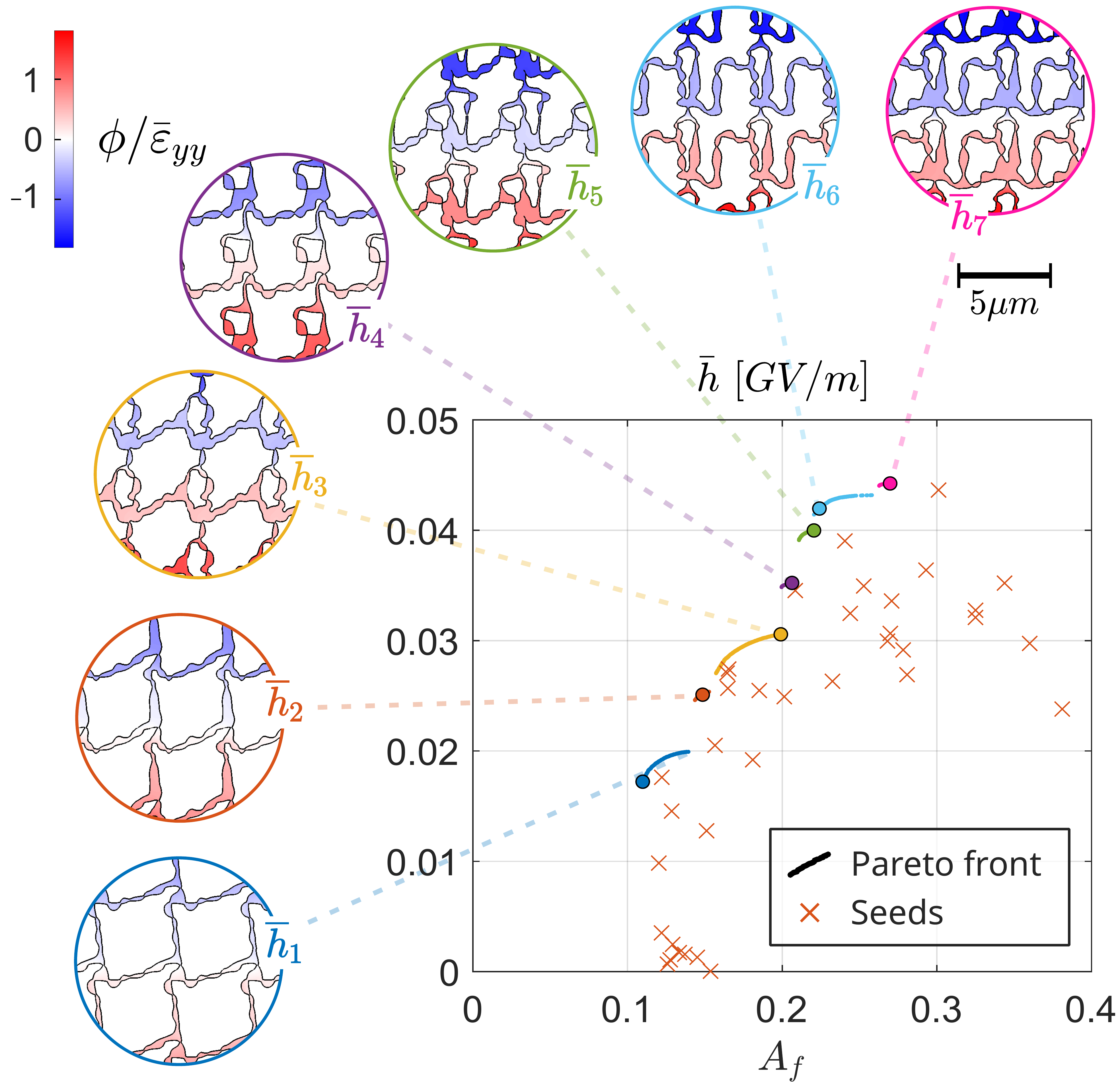}
	\caption{\label{fig:pareto_h}Pareto front of the multi-objective optimization problem for the strain-sensor configuration associated to the equivalent piezoelectric coefficient $\overline{h}$. Each color represents topologically similar microstructures. The insets show selected microstructures, and their colorbars represent normalized quantities for visualization, i.e. the electric potential and macroscopic potential drop are divided by $\mu_\textrm{L}/\kappa$.}
\end{figure}

\begin{figure}  
	\centering
	\includegraphics[width=0.9\textwidth]{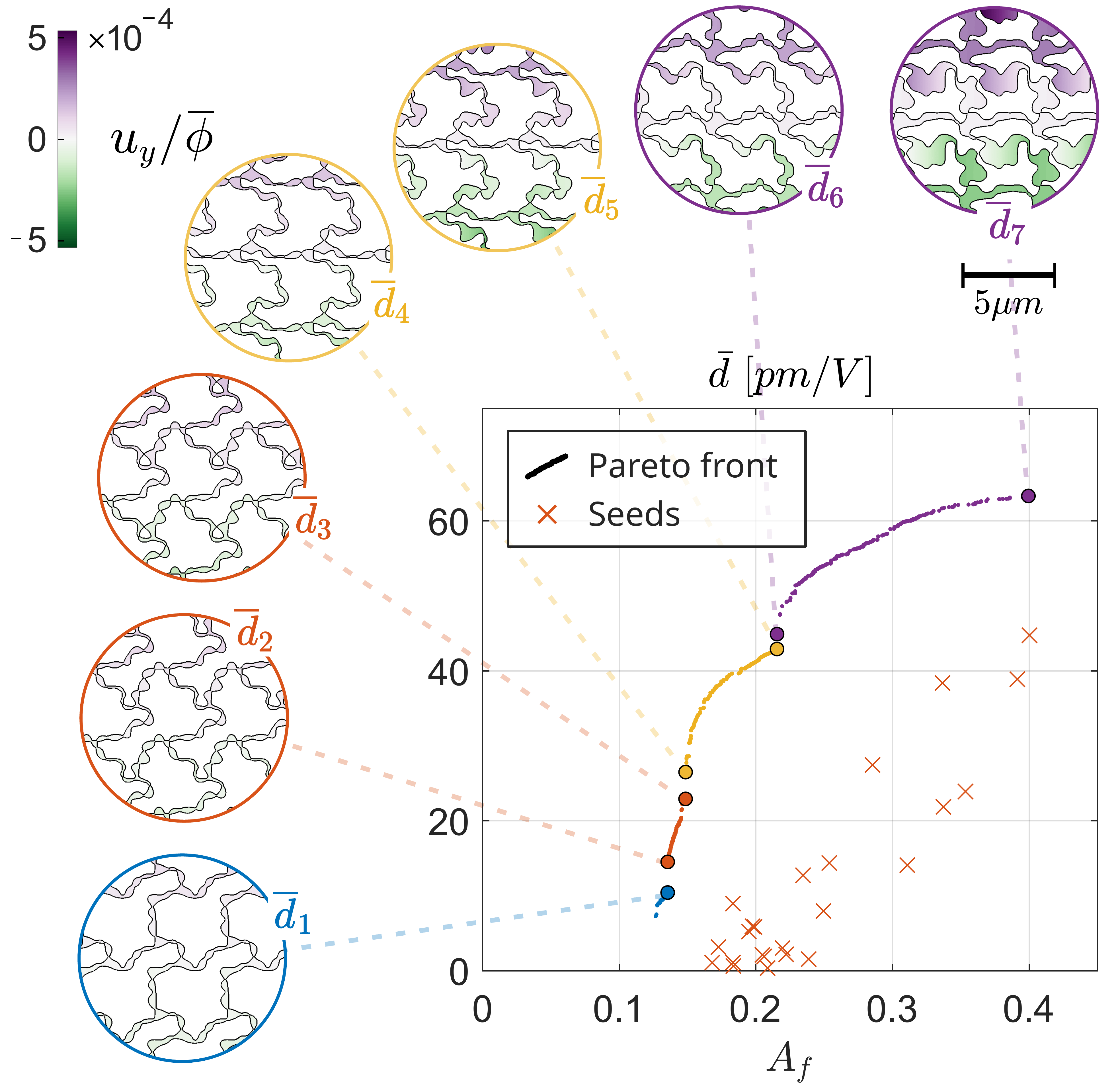}
	\caption{\label{fig:pareto_d} Pareto front of the multi-objective optimization problem for the strain-actuator configuration associated to the equivalent piezoelectric coefficient $\overline{d}$. Each color represents topologically similar microstructures. The insets show selected microstructures, and their colorbars represent normalized quantities for visualization, i.e. the displacement is divided by $L_y$.}
\end{figure}

\begin{figure}  
	\centering
	\includegraphics[width=0.9\textwidth]{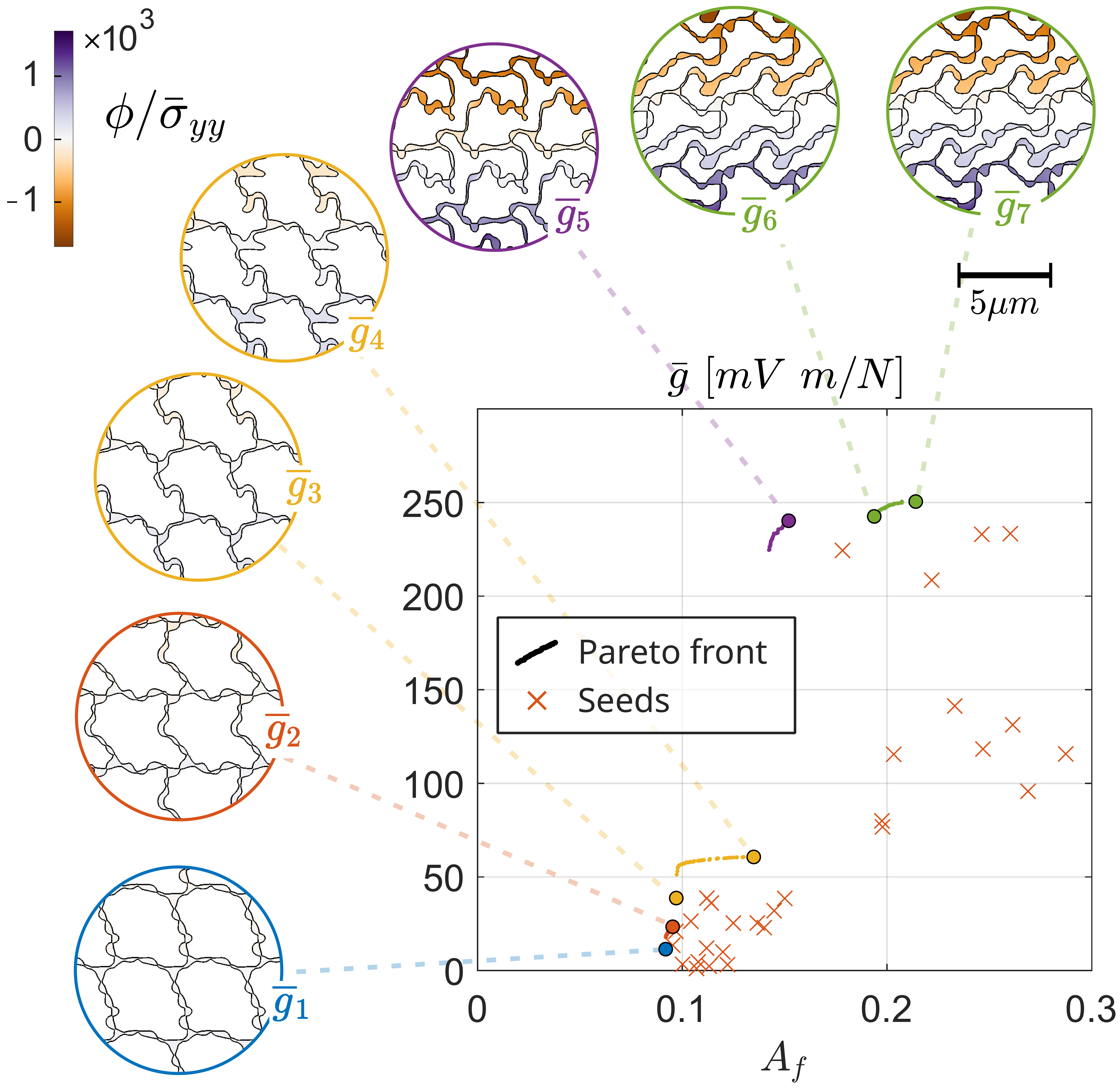}
	\caption{\label{fig:pareto_g}Pareto front of the multi-objective optimization problem for the stress-sensor configuration associated to the equivalent piezoelectric coefficient $\overline{g}$. Each color represents topologically similar microstructures. The insets show selected microstructures, and their colorbars represent normalized quantities for visualization, i.e. the macroscopic stress is divided by $Y$}
\end{figure}

\begin{figure}  
	\centering
	\includegraphics[width=0.9\textwidth]{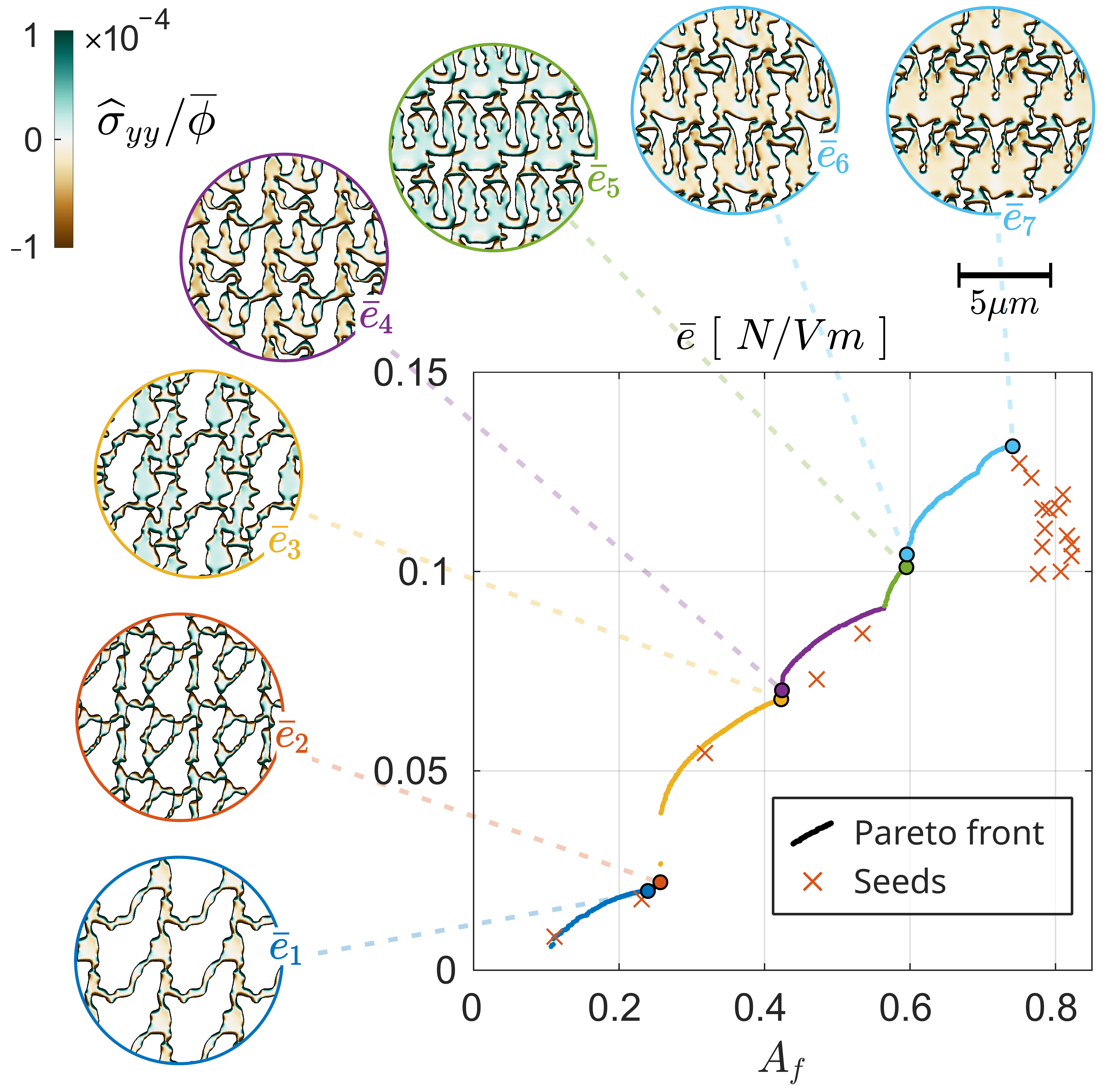}
	\caption{\label{fig:pareto_e}Pareto front of the multi-objective optimization problem for the stress-actuator configuration associated to the equivalent piezoelectric coefficient $\overline{e}$.  Each color represents topologically similar microstructures. The insets show selected microstructures, and their colorbars represent normalized quantities for visualization, i.e.the microscopic Cauchy stress is divided by $Y$.}
\end{figure}

In principle, the interplay between apparent piezoelectricity and area fraction is non-trivial because, on the one hand, higher area fraction uses more material capable of mobilizing flexoelectricity, but on the other hand, slender structures may develop higher field gradients.  
The results of the optimization procedure for each of the four device conditions, are summarized in Fig.~\ref{fig:pareto_h} for the  strain sensor ($\overline{h}$), Fig.~\ref{fig:pareto_d} for the   strain actuator ($\overline{d}$), Fig.~\ref{fig:pareto_g} for the   stress sensor ($\overline{g}$), and Fig.~\ref{fig:pareto_e} for the stress actuator ($\overline{e}$). 
In each case, we also show in the plane of objective functions ($\overline{h}$, $\overline{d}$, $\overline{g}$ or $\overline{e}$ vs.~area fraction $A_f$) the seeds used in the genetic algorithm and the portions of the Pareto front identified by our numerical study. We label with colors branches of the Pareto front characterized by a smooth shape and by similar microstructures. For each device functionality and branch, we plot a pertinent field (electric field, displacement or stress) on representative microstructures.

\subsection{Features of the optimal solutions}
\label{interpret}

We observe that for all four functionalities, our algorithm is able to identify significant portions of the Pareto front, particularly for actuators. The ranges of effective piezoelectric coefficient among Pareto-optimizers varies  between less than 4-fold differences for strain sensors to 25-fold differences for stress sensors, which are very sensitive to area fraction. Our Pareto fronts exhibit relatively narrow ranges of area fraction, between 0.1 and $<0.4$ in all cases, except for stress actuators, where Pareto optimal designs  span a wide range of area fractions, from 0.1 to close to 0.8. For this functionality, more material enables the meta-material to produce a higher stress, but clearly area fraction cannot approach 1 too much since, in this case, the material approaches a centrosymmetric bulk material with vanishing apparent piezoelectricity. In all other cases, our results suggest that optimal performance is achieved at reduced area fraction. 

Regarding the geometry of the optimal meta-materials, our topology optimization algorithm identifies microstructures that share the general features of those proposed heuristically by \cite{mocci2021geometrically}, namely geometric polarization and lattice-like geometries made of slender elements. An exception to this are high-area fraction stress actuators, Fig.~\ref{fig:pareto_e}, which look more like bulk materials with geometrically polarized voids as proposed by \cite{sharma2007possibility}.  We observe that our algorithm identifies unit cells with relatively complex organizations, e.g.~ with three different kinds of voids ($\overline{e}_5$ or $\overline{e}_7$). 

For the strain actuator $\overline{d}$, Fig.~\ref{fig:pareto_d}, the output is the vertical strain at zero stress under an applied electric field. Accordingly, the optimized structures look like scissor mechanisms with non-centrosymmetric sub-units. In this case, we find for the larger area-fraction branches  ($\overline{d}_4$ to $\overline{d}_5$ or $\overline{d}_6$ to $\overline{d}_7$) that increasing feature thickness with similar microstructure shape improves performance. 

Turning to sensor configurations, the goal is to maximize the macroscopic electric field  for a given applied macroscopic strain ($\overline{h}$) or stress ($\overline{g}$). Intuitively, we expect efficient stress sensors to be very compliant along the $y$ direction to maximize mechanical gradients under a given stress and hence produce a strong flexoelectric response. Accordingly,   the optimization procedure leads to apparently floppy structures with horizontally arranged $S$-shaped elements, Fig.~\ref{fig:pareto_g}. In contrast, strain sensors appear stiffer and exhibit vertically arranged elements, which produce higher flexoelectric coupling at fixed applied strain, Fig.~\ref{fig:pareto_h}. In summary,  while \cite{mocci2021geometrically} proposed designs that in broad terms resemble those found here on the basis of fundamental symmetry and heuristic arguments, the unbiased systematic optimization method further produces very different metamaterial designs depending on the sought-after piezoelectric functionality.

\subsection{Macroscopic vs microscopic geometric features}

Besides the overall shapes of the lattice-like optimized metamaterials, they also exhibit oscillatory microscopic geometric features and, in some cases,  small connections in the form of ligaments between bulkier regions. A  similar issue is observed in works using topology optimization based on quadrilateral elements, where a characteristic zigzag pattern is observed on the boundary with a length scale proportional to the mesh size  \citep{jahangiry2017isogeometrical}. Here, because micro-structures are dominated by boundaries, particularly at low area fractions, and the spatial resolution of the geometry is constrained by the computational cost of GAs, this kind of boundary irregularity is particularly evident, with curvy shapes resulting from the higher-order approximation functions.

\begin{figure}  
	\centering
	\subfigure[  ]
	{\includegraphics[width=0.4\textwidth,clip,keepaspectratio,angle=0]{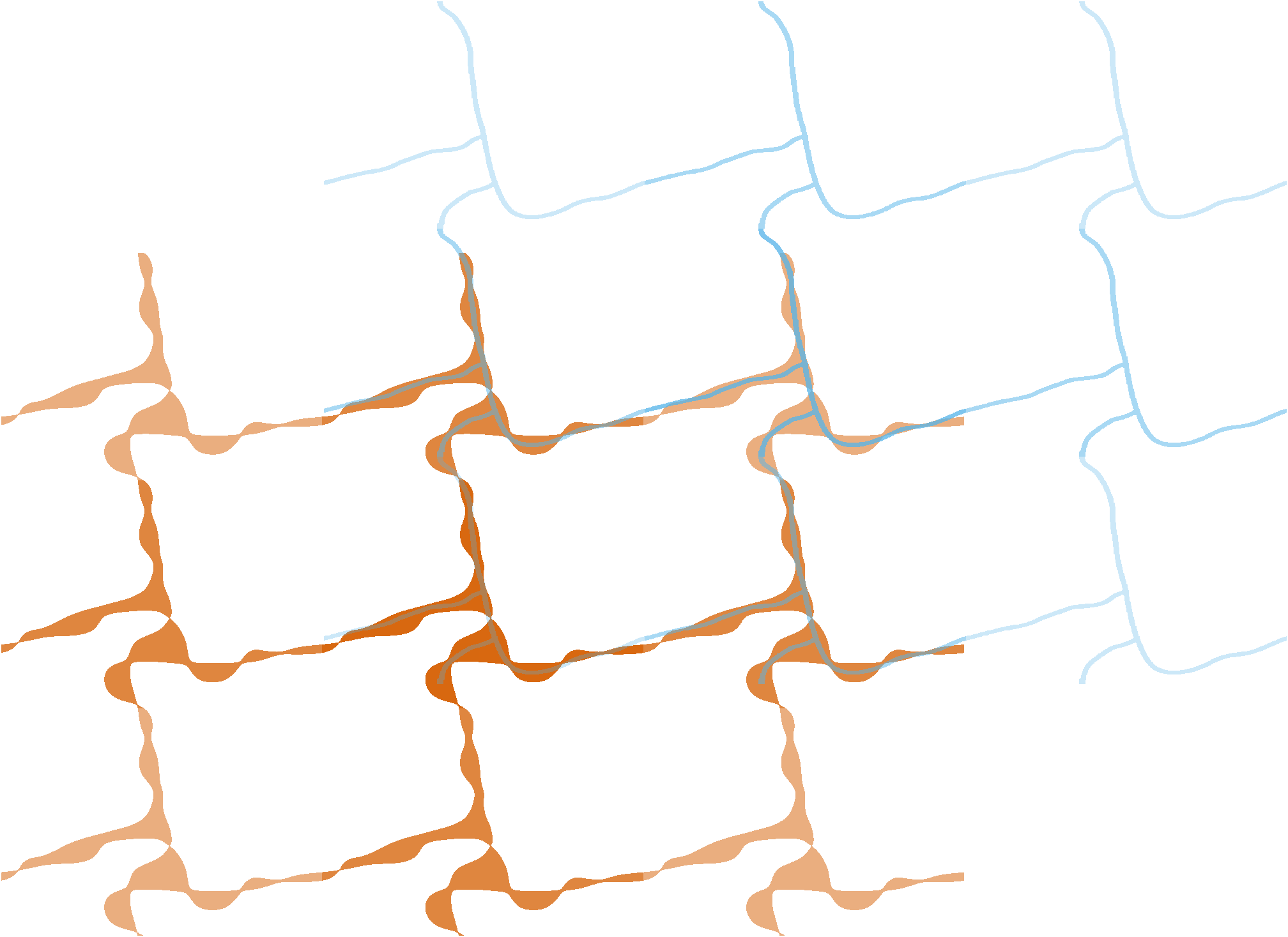}}\;\;\;\;\;
	\subfigure[ ]
	{\includegraphics[width=0.4\textwidth,clip,keepaspectratio,angle=0]{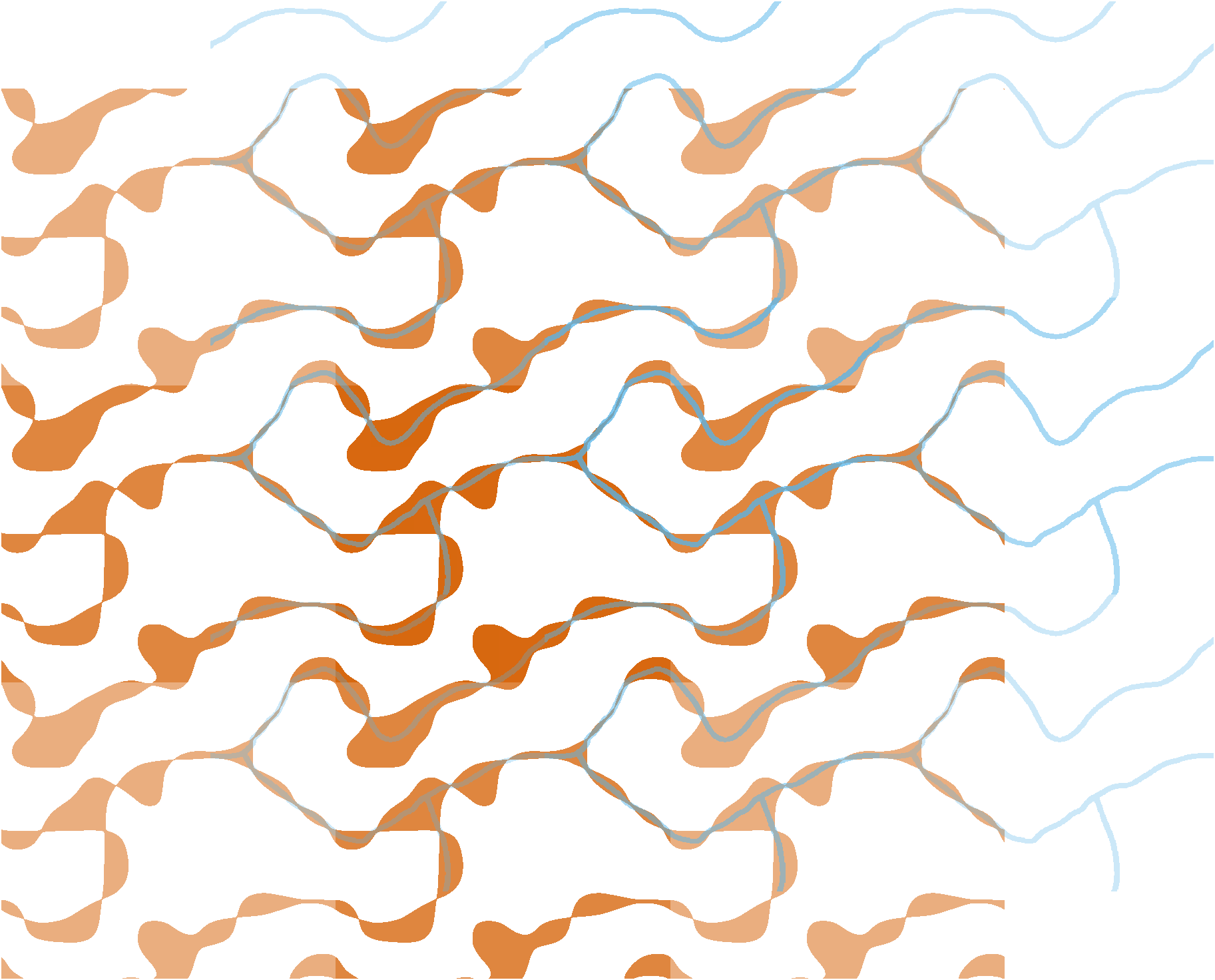}}
	\caption{\label{fig:comp} Starting from the $\overline{h}_1$ structure of Fig.~\ref{fig:pareto_h} (a) and the $\overline{g}_7$ structure of Fig.~\ref{fig:pareto_g} (b) a simpler representation of  their macroscopic topology is extracted by fitting a quadratic B-spline curve to the middle-line of struts.}
\end{figure}

\begin{figure} 
	\centering
	\subfigure[ $t=80$~nm ]
	{\includegraphics[width=0.3\textwidth,clip,keepaspectratio,angle=0]{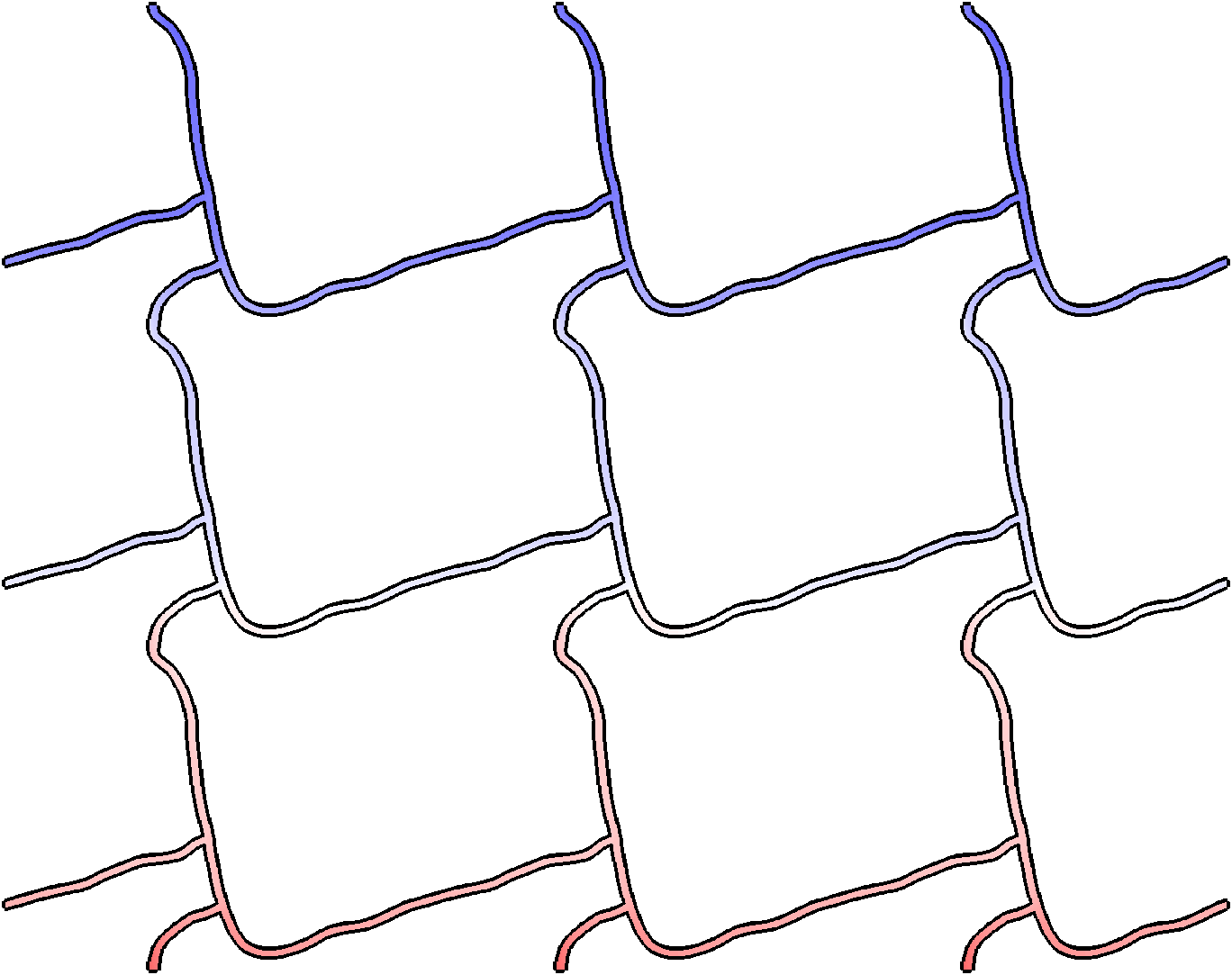}}\;\;\;\;\;
	\subfigure[  $t=160$~nm]
	{\includegraphics[width=0.3\textwidth,clip,keepaspectratio,angle=0]{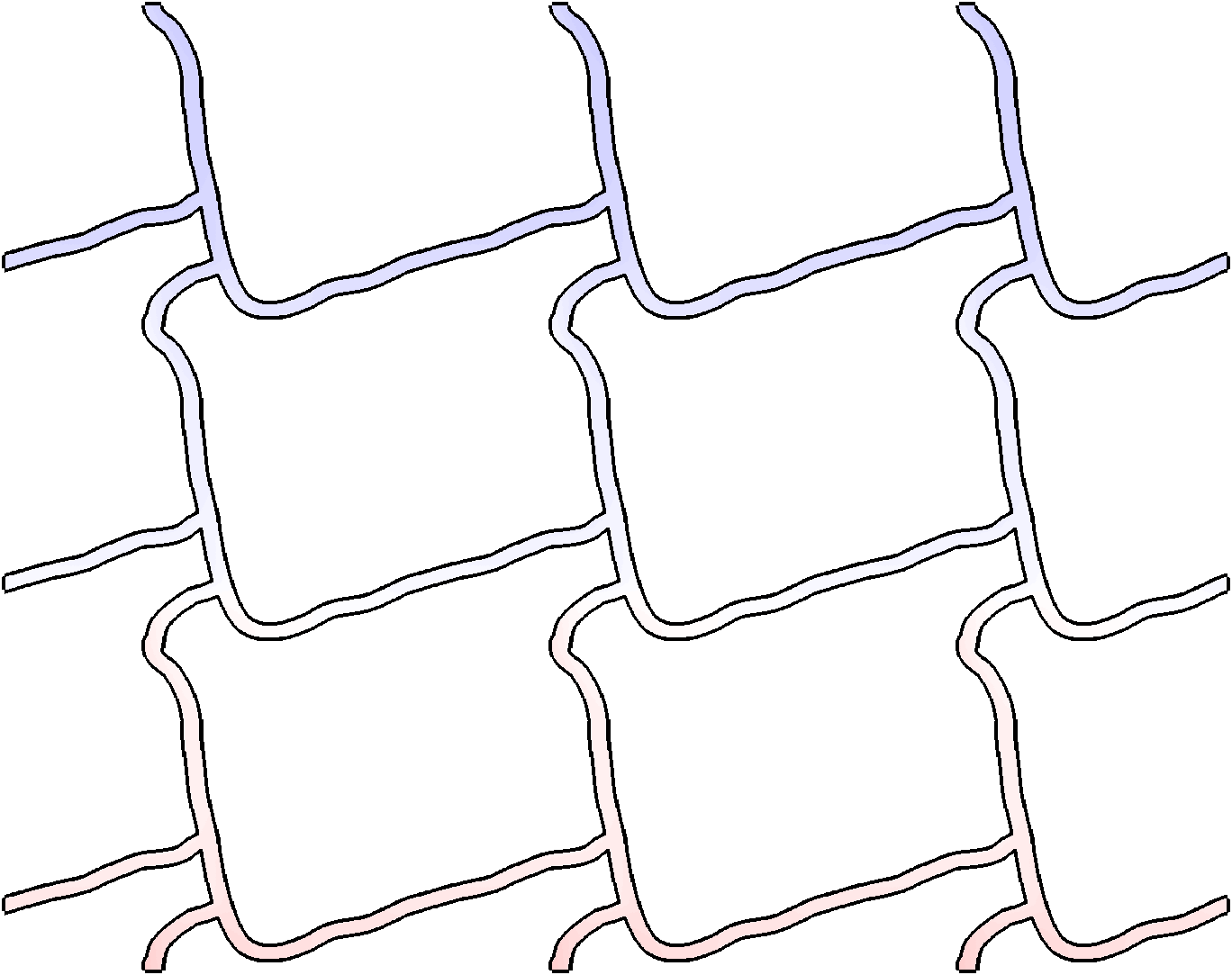}}\;\;\;\;\;
	\subfigure[ $t=240$~nm  ]
	{\includegraphics[width=0.3\textwidth,clip,keepaspectratio,angle=0]{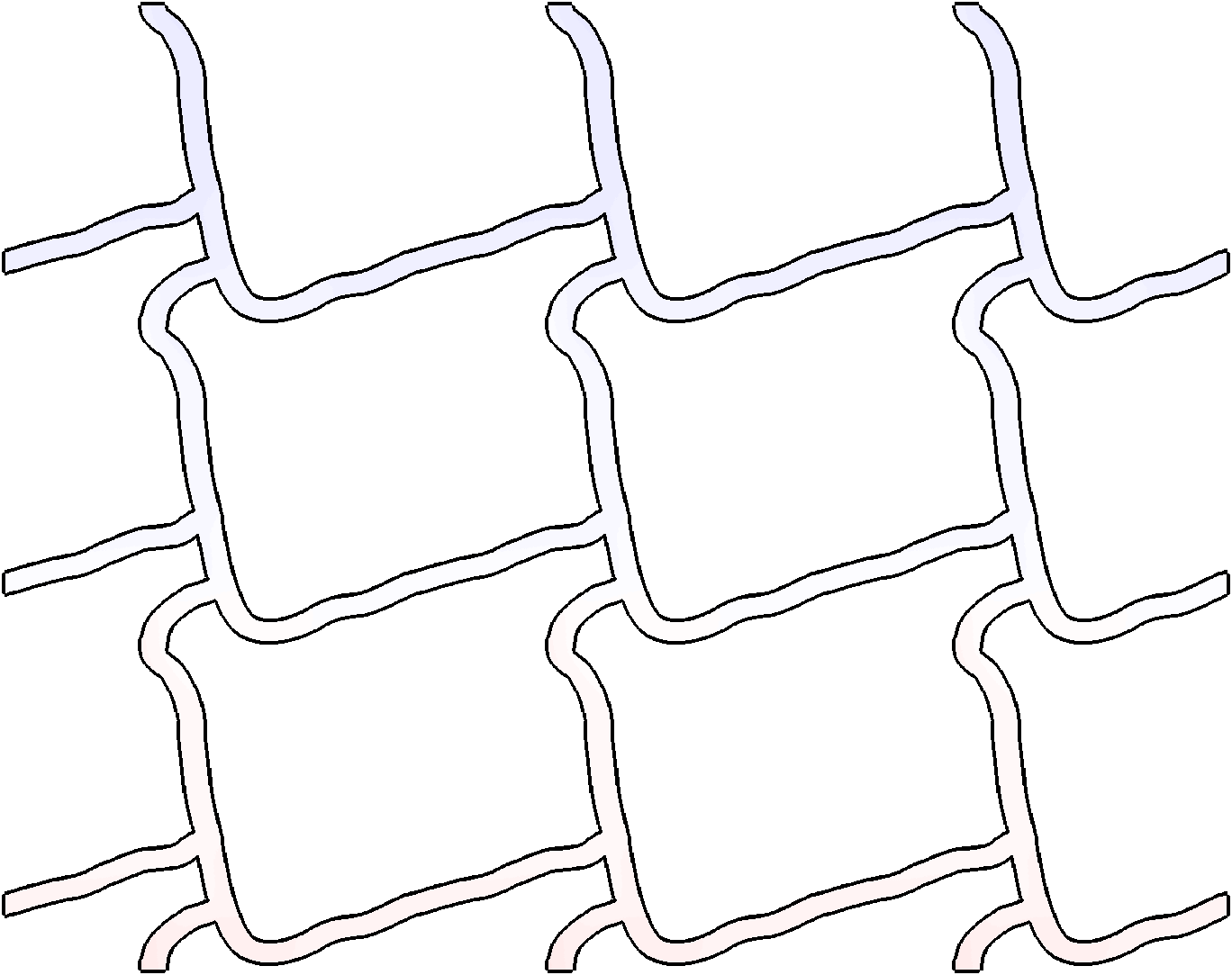}} 
	\caption{ \label{fig:h_appr} Based on the simplified representation of the $\overline{h}_1$ structure of Fig.~\ref{fig:pareto_h} in terms of the middle-line, we consider metamaterials with different constant thickness.}
\end{figure}

\begin{figure}  
	\centering
	\subfigure[$t=80$~nm ]
	{\includegraphics[width=0.3\textwidth,clip,keepaspectratio,angle=0]{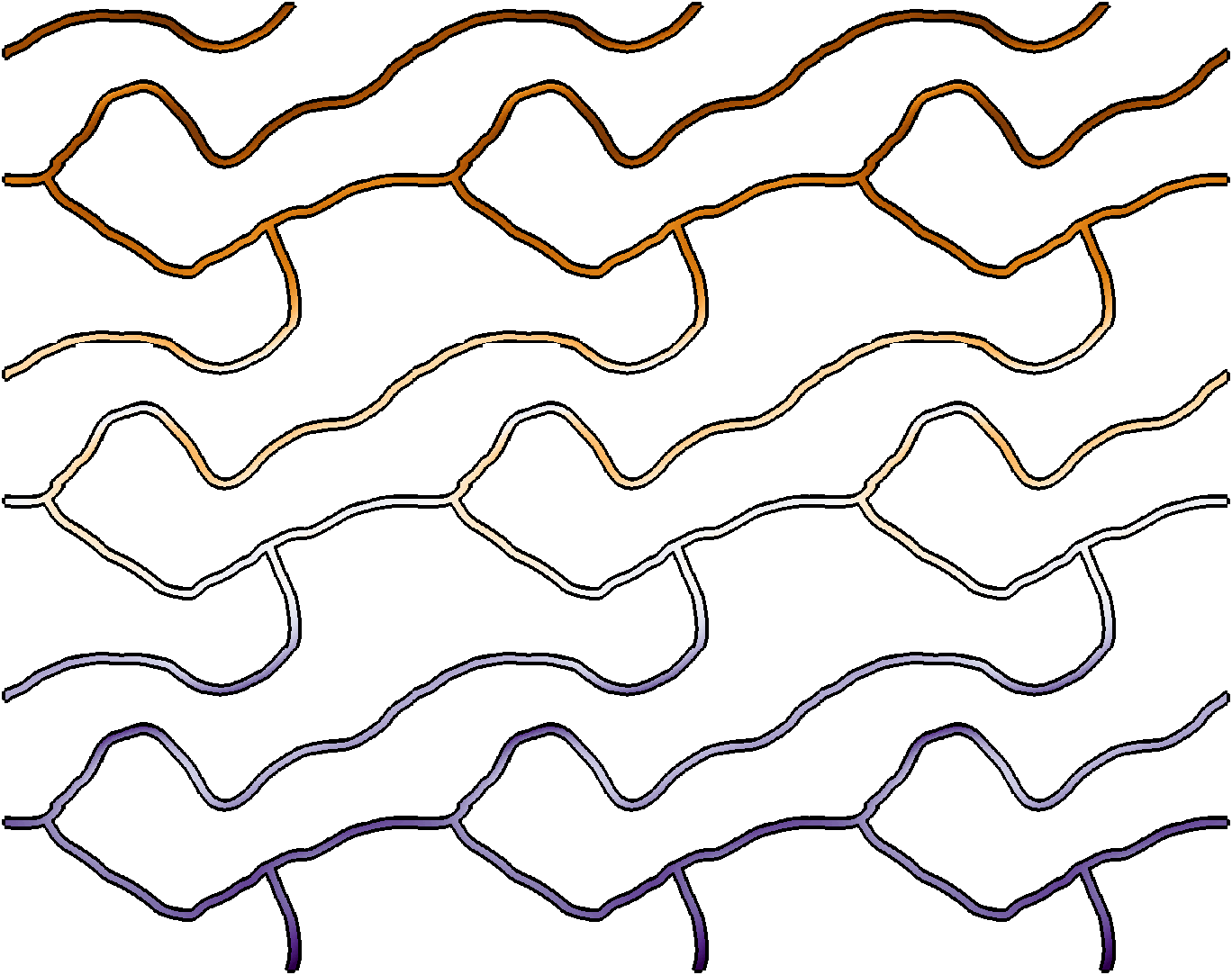}}\;\;\;\;\;
	\subfigure[  $t=160$~nm]
	{\includegraphics[width=0.3\textwidth,clip,keepaspectratio,angle=0]{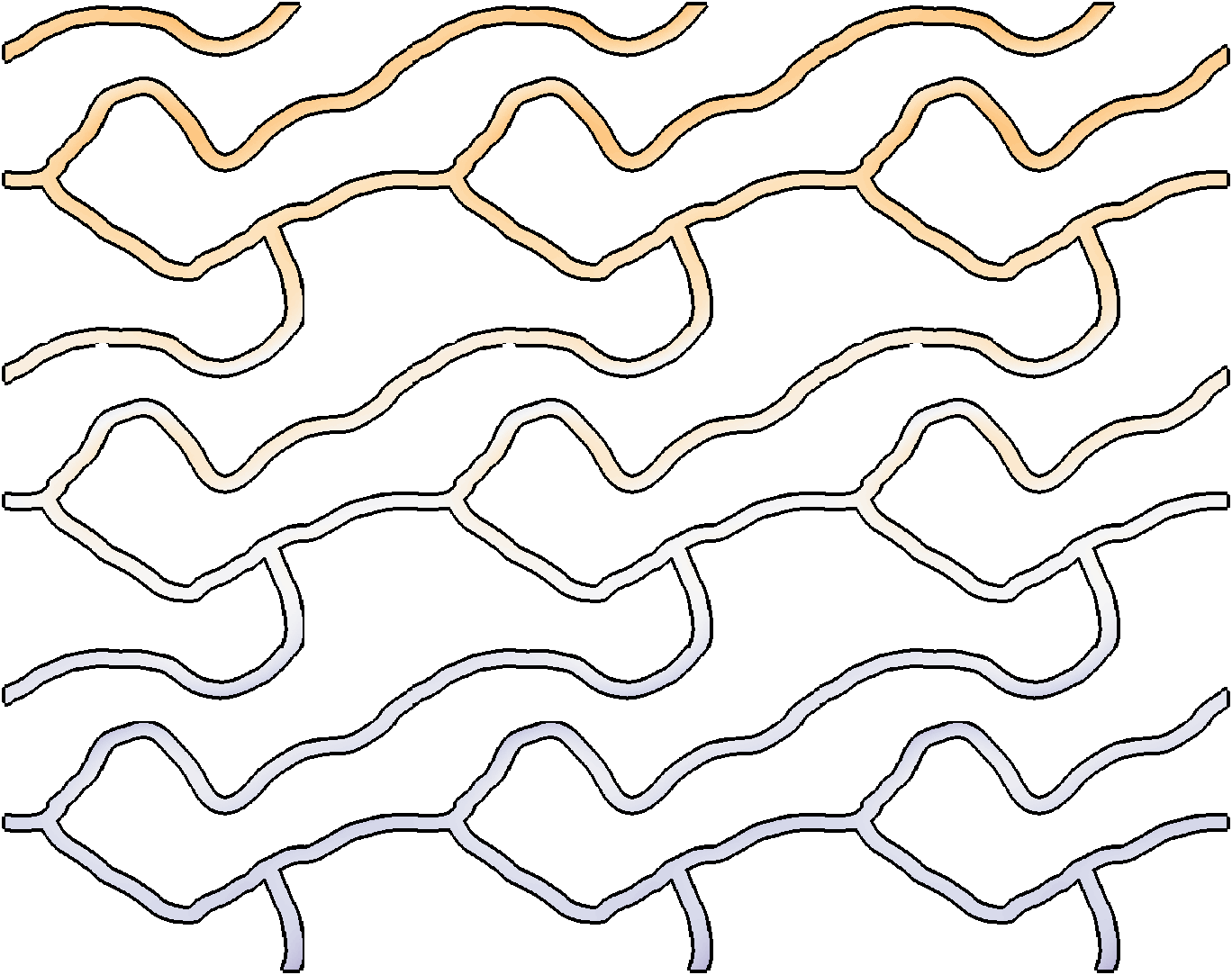}}\;\;\;\;\;
	\subfigure[$t=240$~nm ]
	{\includegraphics[width=0.3\textwidth,clip,keepaspectratio,angle=0]{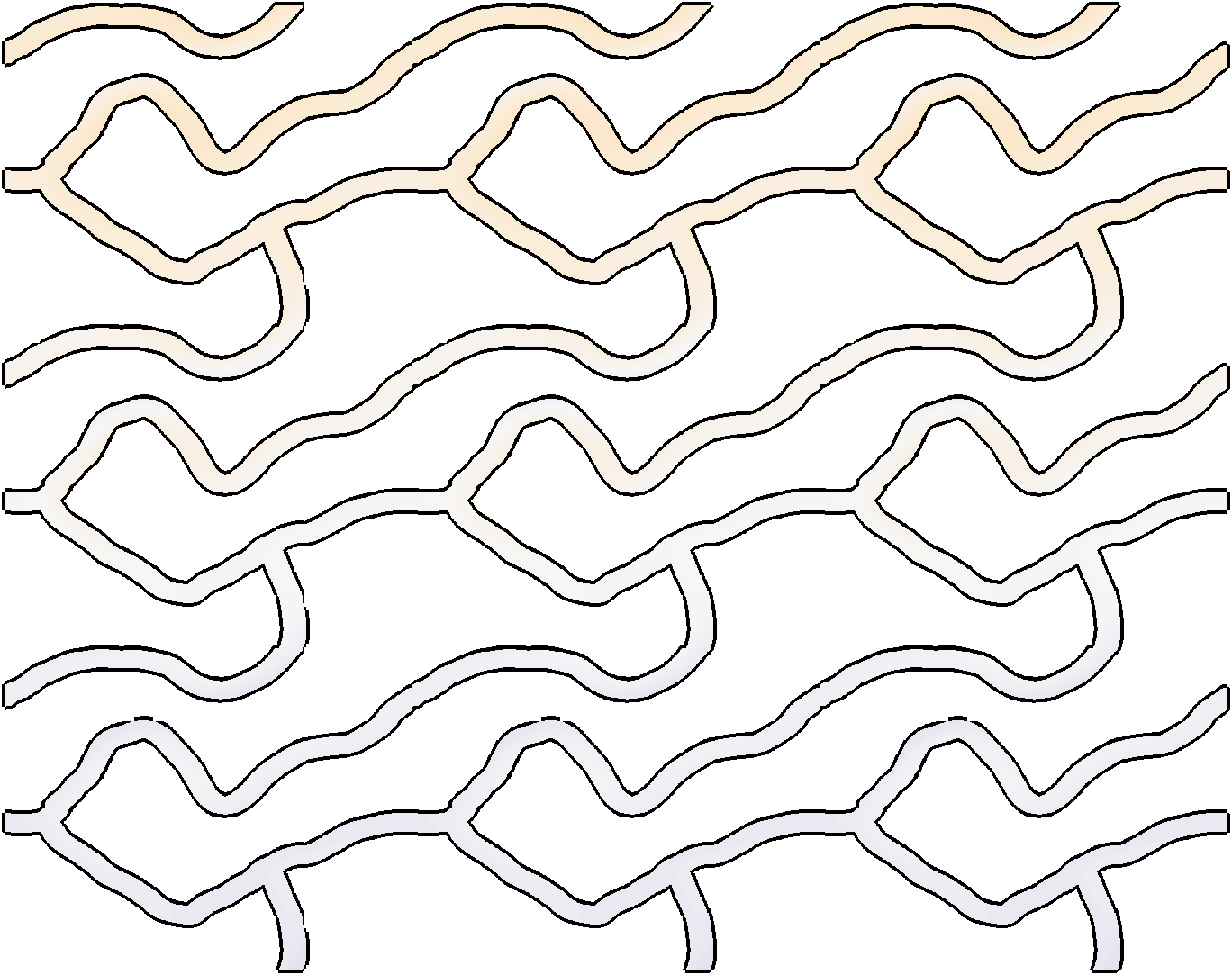}}
	\caption{\label{fig:g_appr} Based on the simplified representation of the $\overline{g}_7$ structure of Fig.~\ref{fig:pareto_g} in terms of the middle-line, we consider metamaterials with different constant thickness.}
\end{figure}

Attenuating these irregularities requires increasing in concert the level of refinement used for geometry and for the discretization of the flexoelectric problem, which as discussed previously, leads to a computationally intractable number of seeds and generations when solved with GA. To assess whether small-scale features have a significant influence on the conclusions of our study, we examine closely two of the optimized structures that have a lattice-like topology, namely the $\overline{h}_1$ structure form Fig.~\ref{fig:pareto_h} and the $\overline{g}_7$ structure from Fig.~\ref{fig:pareto_g}. For each of them, we extract the middle-line of the lattice microstructure and represent it using  quadratic B-spline curves as depicted in Fig.~\ref{fig:comp}. Starting from these curves, we define a family of derived microstructures by thickening by different amounts the middle-line. This is done by thresholding the distance function from the B-spline curves to define a level set representation of the geometry, as  depicted in Fig.~\ref{fig:h_appr} for $\overline{h}_1$ and Fig.~\ref{fig:g_appr} for $\overline{g}_7$.
We consider three thicknesses, 80~nm, 160~nm and 240~nm, and compute the apparent piezoelectric coefficients $\overline{h}$, $\overline{g}$, which are reported in Table~\ref{hap} along with the corresponding area fractions.

These results  confirm the general observations in  \cite{mocci2021geometrically} that $\overline{g}$ (stress sensor) increases as thickness is \textit{uniformly} reduced, whereas $\overline{h}$ (strain sensor) is not very sensitive to  thickness. 
This is not in contradiction with the behavior of the structures of the $\overline{g}$ Pareto front of Fig.~\ref{fig:pareto_g}, where an accumulation of material in some \textit{specific} areas can increase $\overline{g}$. The results in this table also show that the optimal performance depends mainly on the macroscopic shape of the structures, and not on smaller scale waviness of the boundary. This is particularly evident for the case of $\overline{h}$ (strain sensor), which remains nearly the same and close to $\overline{h_1}$ for all thicknesses. 
Although $\overline{g}$  depends on thickness, and hence it is difficult to define a fair comparison with the design  $\overline{g}_7$, our results show that the 80~nm structure leads to $\overline{g}=250.2$~mV~m/N, very close to the optimal $\overline{g}_7=249.8$~mV~m/N f, which involves thin connections that are significantly smaller than 80~nm and a larger area fraction.
In addition, it can be observed that all the structures derived from the optimal  $\overline{h_1}$ consistently deliver higher  $\overline{h}$ (at least by one order of magnitude) than those derived from  $\overline{g_7}$ and, in all cases a lower $\overline{g}$  as compared to $\overline{g_7}$  (Table~\ref{hap}). 

In summary,   our results capture the fundamentally different nature of optimal microstructures for each mode of operation  and demonstrate that the overall nature of optimal microstructures is much more important than microscopic geometric features. These results also show that a simple post-processing step can  be useful, not only to smoothen the boundary and remove unnecessary and spurious microscopic features, but also to generate structures that are easier to manufacture. Finally, these results  suggest an optimization procedure in which the skeleton of the lattice and its thickness are optimized independently. 

Alternative to this post-processing procedure, the  thickness of ligaments can in principle be limited by including in the optimization a constraint based on a stress measure, as for instance in \cite{ortigosa2022computational}. In this work, the thin connections contribute significantly to the flexoelectric objective function, defined as an energy ratio. Our results show that for objective functions based on  macroscopic piezoelectric performance, thin connections do not have such a strong effect. 

 \subsection{Quantitative comparison}

A quantitative comparison with reference values for piezoelectric coefficients of Quartz and PZT ceramics  \citep{kholkin2008piezoelectric}, reported in Table~\ref{tabref}, shows that the apparent piezoelectric coefficients found here are higher than those of Quartz and PZT  for the case of  $\overline{g}$ (stress sensor), and in the same range for the case of  $\overline{d}$ (strain actuator). The values found for  $\overline{e}$ (stress actuator) are competitive only with respect to Quartz, while for   $\overline{h}$ (strain sensor) our metamaterials are significantly less performant than the reference piezoelectric materials. We should note that the piezoelectric properties reported in the table are in absolute terms. When normalized by mass or area fraction, they are much more favorable for our metamaterials. Furthermore, metamaterials with apparent piezoelectricity can in principle be manufactured from any dielectric, can be lead-free and rather insensitive to temperature as opposed to PZT. It is also true that for all four apparent piezoelectric responses, the lowest area fractions may lead to excessively compliant structures. Furthermore, our optimization framework has not considered fabrication constraints \citep{Paulino2016} nor limited maximum stresses \citep{ortigosa2022computational,da_Silva_2021}, strains or electric field, which may result in material damage or electric breakdown.

\begin{table}[]
	\centering
	\caption{\label{hap} Apparent piezoelectric coefficients and area fractions for the optimized structures $\overline{h}_1$ (left) and  $\overline{g}_7$ (right) and their associated simplified versions with uniform thickness.}
	\begin{tabular}[t]{ |c| c| c| c|  }
		\hline
		& $\overline{h}$ & $\overline{g}$ &$A_f$ \\
		 & [GV/m] & [mV~m/N] &  \\  \hline
		Optimum  $\overline{h}_1$ & 0.0171 &3.2 &0.11 \\ \hline
		Beams $t=80$~nm & 0.0147 &40.7 &0.04 \\ \hline
		Beams $t=160$~nm   &0.0141 & 12.4 &0.09 \\ \hline
		Beams $t=240$~nm &0.0133 & 5.1   &0.13   \\ \hline
	\end{tabular}
	\quad  
	\begin{tabular}[t]{ |c| c| c| c|  }
		\hline
		t & $\overline{h}$ & $\overline{g}$ &$A_f$ \\
		nm & [GV/m] & [mV~m/N] &  \\  \hline
		Optimum  $\overline{g}_7$ & 0.0011 &249.2 &0.22 \\ \hline
		Beams $t=80$~nm & 0.0011 &250.2	 &0.07 \\ \hline
		Beams $t=160$~nm  &0.0012 & 79.2 &0.14 \\ \hline
		Beams $t=240$~nm &0.0011 & 31.6   &0.21   \\ \hline
	\end{tabular}
\end{table}

 \begin{table}[]
	\centering
	\caption{\label{tabref}Reference values for piezoelectric coefficients in Quartz and PZT ceramics, compared to the largest apparent piezoelectric coefficients of the microstructures found in this work.}
	\begin{tabular}[t]{ |c| c| c| c|c|  }
		\hline
		& $\overline{h}$ & $\overline{d}$ &$\overline{g}$  & $\overline{e}$ \\  
		&  [GV/m] &     [pm/V]& [mV~m/N] & [N/Vm] \\ \hline
		Quartz \citep{kholkin2008piezoelectric}& 5 & 2.3 &50 &  0.23\\ \hline
		PZT \citep{kholkin2008piezoelectric}&1.4 & 280 &25 & 20.23\\ \hline
		Architected material &0.04 ($\overline{h}_7$) & 63 ($\overline{d}_7$) &250 ($\overline{g}_7$) & 0.13 ($\overline{e}_7$)\\ \hline
	\end{tabular}
\end{table}

\section{Concluding remarks}

We have presented a theoretical and computational framework for the two dimensional topology optimization of flexoelectric metamaterials with apparent piezoelectricity. A key attractive feature of metamaterials as compared to flexoelectric nanostructures is that their multiscale nature enables transcending the nanoscale, where flexoelectricity is significant, to make this effect available for electro-mechanical transduction at  meso- or macroscales. Different actuator  and sensor  functionalities have been optimized, finding numerical approximations of the Pareto fronts with respect to such piezoelectric responses and area minimization. Interestingly, our optimization procedure recovers and improves in an automatic and unsupervised manner the kind of metamaterial RVEs proposed by \cite{mocci2021geometrically} on the basis of physical reasoning, which except for stress actuators consist of a low area-fraction lattice-like geometry with slender elements. However, the identified designs are more complex, highly anisotropic, and markedly different depending on the target functionality, stress/strain sensor/actuator. These results demonstrate a non-trivial structure-property relation in flexoelectric metamaterials with apparent piezoelectricity.

Further emphasizing this idea, \cite{chen2021topology} studied periodic composites, but optimized flexoelectric behavior given a base piezoelectric material (unlike here, where we optimize piezoelectric behavior given a base flexoelectric material), finding non-lattice geometries. Likewise most previous efforts of topology optimization of flexoelectric systems  \citep{nanthakumar2017topology,ghasemi2017level,ghasemi2018multi,zhang2022flexoelectric,lopez2022isogeometric,hamdia2019novel} have focused on structures, not on periodic metamaterials, and have maximized a ratio of electric vs mechanical energies,  not the effective piezoelectric response, leading in all cases to very different designs lacking a lattice structure when the base material is a pure flexoelectric. Taken together, the present and previous works suggest that flexoelectric material design benefits from unbiased optimization techniques and requires a precise specification of the sought-after function. While the physical interpretation of why some structures or metamaterials are optimal is difficult for higher-order theories such as flexoelectricity, the optimal designs obtained here admit a reasonable rationalization, discussed in Section \ref{interpret}.

As a result of our optimization approach, solutions exhibit geometric features at the RVE scale (the shape of the lattice) and at the level of individual struts (a wavy boundary). By removing the latter following a post-processing procedure, we show that the RVE geometry plays a dominant factor. This post-processing  and the nature of our results suggests an optimization procedure where the middle-line of struts and their thickness are the optimization variables. 

Using a good flexoelectric material as base material (BST) for our metamaterials, we obtain apparent piezoelectric coefficients $\overline{h}$,~$\overline{d}$,~$\overline{g}$ and $\overline{e}$ that are  competitive with respect to the reference Quartz and PZT values, illustrating that flexoelectric structures can provide significant apparent piezoelectricity from a much wider and cheaper range of materials. A reference size of $4\upmu m$ has been considered for the unit cell  but, as shown by previous studies, the apparent piezoelectric coefficients can be improved by working at a smaller scale due to the size dependency flexoelectricity. The topology of the optimized structures provide useful insight on how the microscopic flexoelectric behavior is leveraged for each distinct piezoelectric functionality, which can inspire simpler and more manufacturable structures. Finally, a significant improvement is expected by 3D optimization of flexoelectric metamaterials, the topic of future research.

\section*{Acknowledgements}
This work was supported by the European Research Council (StG-679451 to I.A.), the Spanish Ministry of Economy and Competitiveness (RTI2018-101662-B-I00) and through the ``Severo Ochoa Programme for Centres of Excellence in R\&D'' (CEX2018-000797-S), and the Generalitat de Catalunya (ICREA Academia award for excellence in research to I.A., and Grant No. 2017-SGR-1278). D.C. acknowledges the support of the Spanish Ministry of Universities through the Margarita Salas fellowship (European Union-NextGenerationEU).

\section*{CRediT authorship contribution statement}
\textbf{F. Greco}: Conceptualization, Methodology, Software, Validation, Formal analysis, Data curation, Visualization, Writing – original draft.  
\textbf{D. Codony}:  Conceptualization, Methodology, Software, Validation, Formal analysis, Data curation, Visualization, Writing – original draft.
\textbf{H. Mohammadi}: Conceptualization, Methodology, Writing – original draft.
\textbf{S. Fernandez}: Conceptualization, Methodology, Supervision.
\textbf{ I. Arias}: Conceptualization, Methodology, Supervision, Writing – review and editing, Resources, Project administration, Funding acquisition.


\end{document}